\documentclass{ws-ijmpa}
\usepackage[super,compress]{cite}

\usepackage{amsrefs, amsmath}
\usepackage{amsfonts, mathrsfs}
\usepackage{amssymb, url}
\usepackage{amscd, mathtools}
\usepackage{cleveref, multirow} 
\usepackage{tikz,xypic,tikz-cd}

\numberwithin{equation}{section}
\usetikzlibrary{decorations.pathreplacing,angles,quotes}

\crefname{equation}{Eq.}{Eqs.}
\crefname{eqnarray}{Eq.}{Eqs.}
\crefname{algo}{Algorithm}{Algorithms}
\crefname{conj}{Conjecture}{Conjectures}
\crefname{lem}{Lemma}{Lemmas}
\crefname{claim}{Claim}{Claims}
\crefname{rmk}{Remark}{Remarks}
\crefname{prop}{Proposition}{Propositions}
\crefname{section}{Section}{Sections}
\crefname{appendix}{Appendix}{Appendices}
\crefname{cor}{Corollary}{Corollaries}
\crefname{figure}{Figure}{Figures}
\crefname{table}{Table}{Tables}
\crefname{example}{Example}{Examples}
\crefname{prob}{Problem}{Problems}
\crefname{assm}{Assumption}{Assumptions}
\crefname{defn}{Definition}{Definitions}

\newcommand{\de}{{\partial}}

\newcommand{\rd}{\mathrm{d}}
\newcommand{\ri}{\mathrm{i}}
\newcommand{\re}{\mathrm{e}}

\newcommand{\bbA}{\mathbb{A}}
\newcommand{\bbO}{\mathbb{O}}
\newcommand{\bbN}{\mathbb{N}}

\newcommand{\bbZ}{\mathbb{Z}}
\newcommand{\bbR}{\mathbb{R}}
\newcommand{\bbC}{\mathbb{C}}
\newcommand{\bbP}{\mathbb{P}}
\newcommand{\bbF}{\mathbb{F}}
\newcommand{\bbQ}{\mathbb{Q}}
\newcommand{\bbT}{\mathbb{T}}

\def\bary{\begin{array}} 
\def\eary{\end{array}} 
\def\ben{\begin{enumerate}} 
\def\een{\end{enumerate}}
\def\bit{\begin{itemize}} 
\def\eit{\end{itemize}}
\def\nn{\nonumber} 

\newcommand{\cY}{\mathcal{Y}}

\newcommand{\cO}{\mathcal{O}}

\newcommand{\cN}{\mathcal{N}}
\newcommand{\cW}{\mathcal{W}}

\newcommand{\cF}{\mathcal{F}}

\newcommand{\cX}{\mathcal{X}}

\newcommand{\cM}{\mathcal M}

\def\beq{\begin{equation}}                     %
\def\eeq{\end{equation}}                       %
\def\bea{\begin{eqnarray}}                     
\def\eea{\end{eqnarray}}
\def\bary{\begin{array}} 
\def\eary{\end{array}} 
\def\ben{\begin{enumerate}} 
\def\een{\end{enumerate}}
\def\bit{\begin{itemize}} 
\def\eit{\end{itemize}}
\def\nn{\nonumber} 
\def\de {\partial}

\newcommand{\GITl}[1]{\backslash \!\! \backslash _{\kern-.2em #1 \kern0.1em}}
\newcommand{\GIT}[1]{/\!\!/_{\kern-.2em #1 \kern0.1em}}

\renewcommand{\l}{\left}
\renewcommand{\r}{\right}

\newcommand{\rank}{\operatorname{rank}}

\newcommand{\qbinom}{\genfrac{[}{]}{0pt}{}}

\def\bred{\begin{color}{red}}
\def\ered{\end{color}}

\def\bes{\begin{subequations}}
\def\ees{\end{subequations}}

\begin{document}

%
\catchline{}{}{}{}{}
%

\title{Enumerative geometry of surfaces and topological strings
}

\author{Andrea Brini
}

\address{School of Mathematics and Statistics, University of Sheffield\\  Hounsfield Road,
Sheffield, S3 7RH, United Kingdom\footnote{On leave from CNRS, DR 13, Montpellier, France}\\
a.brini@sheffield.ac.uk}

\maketitle

\begin{history}
\received{Day Month Year}
\revised{Day Month Year}
\end{history}

\begin{abstract}

This survey covers recent developments on the geometry and physics of Looijenga pairs, namely pairs $(X,D)$ with $X$ a complex algebraic surface and $D$ a singular anticanonical divisor in it. I will describe a surprising web of correspondences linking together several a priori distant classes of enumerative invariants associated to $(X,D)$, including the log Gromov--Witten invariants of the pair, the Gromov--Witten invariants of an associated higher dimensional Calabi--Yau variety, the open Gromov--Witten invariants of certain special Lagrangians in toric Calabi--Yau threefolds, the Donaldson--Thomas theory of a class of symmetric quivers, and certain open and closed BPS-type invariants. I will also discuss how these correspondences can be effectively used to provide a complete closed-form solution to the calculation of all these invariants.

\end{abstract}

\ccode{PACS numbers:}


\section{Overview}	

\subsection{Ancient geometry and modern physics}
\label{sec:over}
Enumerative geometry -- the count of geometric configurations, satisfying a suitable set of conditions, inside a given shape -- is a venerable subfield of Mathematics, with roots dating back to Greek Antiquity. Some example questions, in chronological order, are:

\begin{description}
\item[Q1] What is the maximal number of circles tangent to three given circles in the plane? \\
\item[Q2] How many lines are there on a smooth complex cubic surface? 
\item[Q3] How many rational complex plane curves of given degree pass through a suitable number of points?
\item[Q4] How many curves of given degree and genus exist on a Calabi--Yau threefold?
\end{description}
 
While similar in flavor, these problem are dramatically different in their sophistication.  Question~{\bf Q1} was formulated in lost work of Apollonius of Perga (III-II~BC), with an account of the solution known from a report of Pappus of Alexandria (II~AD): its answer (eight) was given using compass and straightedge by Vi\`ete in his {\it Apollonius Gallus} at the end of the XVII century. The answer to Question~{\bf Q2} (twenty-seven) was provided by Cayley--Salmon in the mid XIX century; a proof can be formulated as either an exercise in Schubert calculus or in the geometry of the blow-up of the plane at six points.  Question~{\bf Q3} is classical and equal to one for degree one (lines through two non-coincident points) and two (conics through five points, no three collinear); however it gets significantly harder for higher degrees, with the state-of-the-art stopping at degree five at least until the early 90's of the XX century, when a remarkable formula using non-classical methods was provided by Kontsevich from the geometry of moduli of space of curves. Question~{\bf Q4}, finally, is one of the central problems in the modern enumerative theory of curves, and possibly the most famous also due to its relevance in physics. Calabi--Yau threefolds are supersymmetric backgrounds for four-dimensional type II string compactifications, and the count of curves in them simultaneously computes F-terms involving the Weyl supermultiplet in the four dimensional low-energy effective field theory (the curve counts being identified with worldsheet instanton contributions to the effective action as Gromov--Witten invariants \cite{Antoniadis:1993ze, Bershadsky:1993cx}) and encodes the degeneracy of BPS states in the four-dimensional theory (the counts being recast in the form of BPS degeneracies of D2-branes wrapped on curves \cite{Gopakumar:1998jq}). \\

As these examples show, despite the venerable past of the field and the deceptively innocent-looking flavor of its most basic problems, the field of enumerative (algebraic) geometry has a present inextricably linked to developments in
Mathematical Physics. In particular, the interaction with String Theory has sent shockwaves through the subject, giving both unexpected new perspectives and a remarkably powerful, physics-motivated toolkit to tackle several traditionally hard questions in the field: generating functions of curve counting invariants give rise to $\tau$-functions of some very special infinite-dimensional dynamical systems \cite{Witten:1990hr,Kontsevich:1992ti,dubrovin:2001}, they are special (quasi)-modular forms \cite{Aganagic:2006wq}, and they provide a surprising way to compute sophisticated topological invariants of 3-manifolds and links in them \cite{Gopakumar:1998ki}. \\

The purpose of this survey is to cover a web of novel correspondences adding to this list. It is related to curve counts in complex dimension {\it two}, and is motivated by the physics of topological strings in complex dimension {\it three}. 

\subsection{Two-dimensional geometry}

The central objects of study in this survey will be the enumerative geometry of curves inside \emph{Looijenga pairs}: these are pairs $(X,D)$ with
%
$X$ a complex surface, and $D$ a singular anti-canonical divisor in it. 
%
We will be interested in particular in several classes of enumerative invariants of the \emph{pair} $(X,D)$, that ``feel'' both the geometry of $X$ and that of the divisor $D$. 

\begin{example}
The running example of a Looijenga pair for us will be given by $X=\bbC \bbP^2$, the complex projective plane, and $D=H \cup Q$, with $Q$ a plane conic and $H$ a line not tangent to it; see Figure~\ref{fig:p2lc}.
\begin{figure}
\centering
\includegraphics[scale=.5]{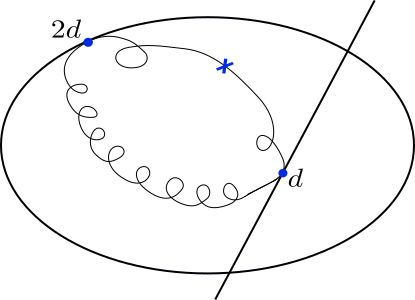}
\caption{A real section of the Looijenga pair geometry $(\bbC \bbP^2, H \cup Q)$. The curly line depicts a degree-$d$ curve, passing through a general point in the plane marked by the blue X, and with maximal contact order at the line $H$ and the quadric $Q$.}
\label{fig:p2lc}
\end{figure}
A special class of enumerative invariants sensitive to the geometry of the pair $(\bbC \bbP^2, H \cup Q)$ is the following. Let $C_d$ be a plane curve of degree $d > 0$: by definition, $C_d$ is the zero locus of a non-zero degree~$d$ homogeneous polynomial $P_d(x,y,z) \in H^0(\bbC \bbP^2, \cO(d))$. The equation defining $C_d$ is determined by the coefficients of the polynomial $P_d$ up to overall scaling: this gives $\binom{d+2}{2}-1$ degrees of freedom in specifying $C_d$. By B\'ezout's theorem, the resulting curve will intersect the line $H$ and the quadric $Q$ at, respectively, $d$ and $2d$ generically distinct points. It will furthermore have genus $g(C_d)=\binom{d-1}{2}$ by the degree-genus formula, generically equal to its topological genus. Now we can look at the maximally non-generic case when the following two conditions are realized:
\begin{description}
\item[maximal tangency:] the intersection points with the line and the conic coalesce into  \emph{single} (unspecified) points $p_H\in H$ (resp. $p_Q \in Q$) on each of them, with maximal contact orders equal to $d$ (resp. $2d$);
\item[rationality:] the curve $C_d$ has vanishing geometric genus.
\end{description}
Imposing that $d$ points come together on the line, and $2d$ on the conic, gives $(d-1) + (2d -1)=3d-2$ constraints on the coefficients of $P_d$. The vanishing genus condition further imposes $\binom{d-1}{2}$ (non-linear) constraints on the coefficients of $P_d$, from the degree-genus formula. The variety of curves satisfying both conditions will then have dimension $$\frac{d(d+3)}{2} - (d-1)-(2d-1) - \frac{d(d-1)}{2}=1.$$ If we further ask that the curve $C_d$ pass through a given (generic) point in the plane, we are left with a zero-dimensional family -- a finite number $\mathfrak{n}_d$ -- of configurations satisfying all constraints, and we can then ask what that number is when these conditions are imposed generically:\\
%
\beq
\mathfrak{n}_d 
 =  \# \left\{ \bary{c} \hbox{ rational plane degree-}d\hbox{ curves through a generic point } \\
 \hbox{and maximally tangent at a line and a conic} \eary \right\}.
\label{eq:ndintro}
\eeq
%
\label{ex:p2lcintro}
\vspace{.5cm}\end{example}

\subsection{Three-dimensional physics}

The classical-looking surface counts of the previous section turn out to be surpisingly related to certain open amplitudes in a supersymmetry-protected, topological subsector of a type IIA non-gravitational string compactification, possibly including D-branes. Counts of curves in a target K\"ahler manifold $Y$ are known to arise in physics as instanton numbers of topologically A-twisted 2d topological $\sigma$-models coupled to worldsheet gravity: a critical example is given by the case in which the target space is a Ricci-flat manifold of complex dimension three, where the partition function is well-defined and non-trivial at all genera. Furthermore, it is possible to place natural Dirichlet boundary conditions preserving half of the worldsheet supersymmetry, which (in absence of a B-field and the gauge field of the brane) amounts to requiring that the boundary of the worldhseet is constrained on a Lagrangian submanifold $L \subset Y$; counts of curves with boundary on the Lagrangian are then physically realized as Euler numbers of moduli spaces of open worldsheet instantons to the Lagrangian pair $(Y,L)$. 

As ramification conditions on a point, such as the one that encodes the tangency conditions in Example~\ref{ex:p2lcintro}, are closely modeled on open boundary conditions around a circle bounding a small disk around the point, one might wonder whether there's a modern physics story behind the type of classical-looking curve counts inside a complex Fano surface in \eqref{eq:ndintro}. Two obvious obstructions to establishing a relation between Looijenga pairs $(X,D)$ and open string counts for  (special) Lagrangians inside Calabi--Yau threefolds are the mismatch in dimension, and the lack of a Calabi--Yau condition on the surface side. One of the salient points of this survey is to describe how such a surprising relation may be realized:  in particular, under relatively lax conditions, a nef Looijenga pair $(X,D)$ will have an associated special Lagrangian pair $(Y,L)$ with $Y$ a toric Calabi--Yau threefold, and $L$ a (framed) Lagrangian in it, whose open topological A-model amplitudes of $(Y,L)$ return curve counts on the surface pair $(X,D)$, both at genus zero and at higher genus.

\begin{example}
The basic example to have in mind is $Y=\bbC^3$, the three-dimensional complex affine space, and $L$ an Aganagic--Vafa brane at some framing $f$. The relevant open string geometry is described by the toric diagram in Figure~\ref{fig:fv}.
\begin{figure}
\centering
\includegraphics[scale=1.5]{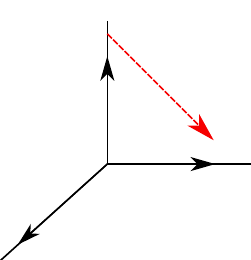}
\caption{The toric web diagram of the framed vertex $(\bbC^3, L)$ at framing one.}
\label{fig:fv}
\end{figure}
Since $(Y,L)$ is a toric Lagrangian pair, it has an open $B$-model mirror described by a curve in $\bbC^* \times \bbC^*$ \cite{Hori:2000ck,Aganagic:2000gs,Brini:2011ij, Fang:2011qd}. In particular, the topological A-model disk amplitude $y(x)$ on $(Y,L)$, as a function of the open string modulus $x \in H^1(L, \bbC)$, satisfies the trigonometric equation 
\beq
1+\re^{x-f y(x)} + \re^{y(x)}=0
\eeq
where $f \in \bbZ$ is the framing of $L$. We will see that for $f=1$, the Taylor coefficients of the disk amplitude $y(x)$ encode in a precise way the solution to the counting problem $\mathfrak{n}_d$ in the previous Section.

\vspace{.5cm}\end{example}

\subsection{The correspondences: key take-aways}

The open topological A-model on a toric CY3 is an extremely well studied subject in both geometry and physics, carrying with itself a spectacular array of theoretical angles and solution methods alike. On the flip side, curve counts in Looijenga pairs are a relatively younger subject, with an ever growing list of open questions: the relation to the topological string on a threefold and its different physical incarnations  becomes then a remarkably powerful means to address them. The three main upshots are as follows:

\ben
\item  there are \emph{several classes of enumerative invariants} of curves attached to the datum of a Looijenga pair $(X,D)$. These include the log~Gromov--Witten invariants of the pair, its local invariants, the open Gromov--Witten theory of an associated toric Lagrangian pair $(Y,L)$, the Donaldson--Thomas invariants of an associated quiver, and a class of open and closed Gopakumar--Vafa-type invariants;
\item although their significance and intepretation varies considerably, these invariants are nonetheless \emph{all related};
\item and furthermore, the problem of computing them is \emph{closed-form solvable}: there is a non-recursive master formula determining them all. 
\een

\begin{figure}
\centering
\includegraphics{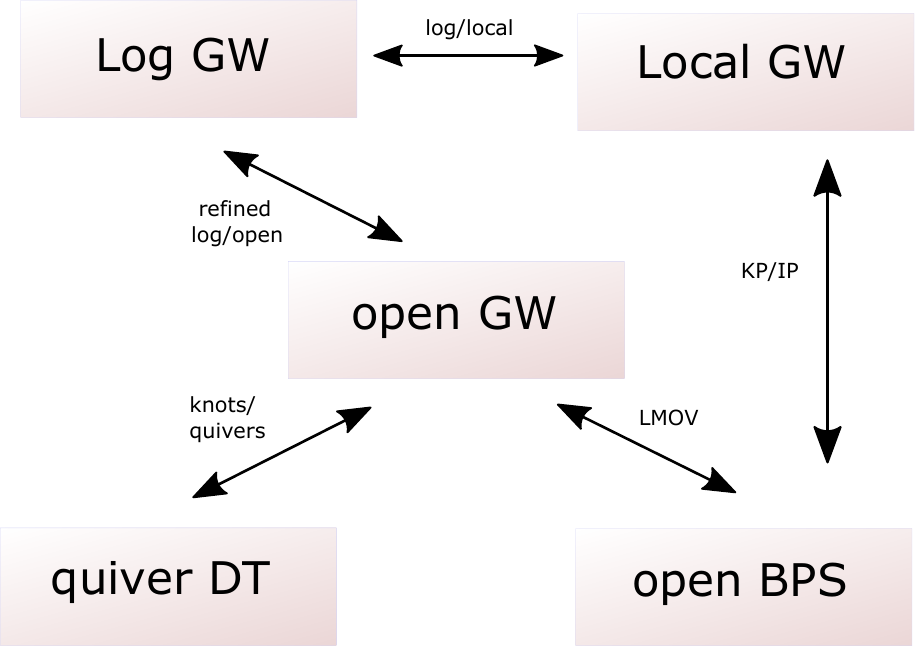}
\caption{The web of correspondences between invariants of Looijenga pairs.}
\label{fig:web}
\end{figure}

The resulting web of correspondences is depicted in Figure~\ref{fig:web}. One practical upshot of having these correspondences in place, aside for their intrinsic interest, is that often solution methods are scarce but for a single type of invariants: we can then use the relations in Figure~\ref{fig:web} to provide closed-form solutions for {\it all} of them, in a single go.\\

This survey is conceived as a gentle introduction and broad overview on the content of \cite{Bousseau:2019bwv,Bousseau:2020fus, Bousseau:2020ryp,Brini:2022htq}, to which the reader is encouraged to turn for additional details. As these papers were specifically targeted to an algebro-geometric audience, the presentation here has an additional slant towards physics, in the hope that a mathematically-minded string theorist can follow through the development of the arguments with relative ease. Although we will barely be able to scratch the surface of some of the concepts introduced (for example log~Gromov--Witten theory and its relation to the Gross--Siebert program), and there is essentially no new material presented here that was not contained in the references above, we will nonetheless try to be reasonably self-contained, and illustrate the correspondences in Figure~\ref{fig:web} with many detailed examples along the way. We should mention that after the appearance of \cite{Bousseau:2019bwv,Bousseau:2020fus, Bousseau:2020ryp} several works have appeared touching on topics closely related to those of this survey; a non-exhaustive list is \cite{BNTY,Graefnitz:2020dlr,Bardwell-Evans:2021tsz,Nabijou:2022utj,Grafnitz:2022wks,Do:2022knz}.

The presentation will be structured as follows. We first give in Section~\ref{sec:gwintro} a quick review of the geometry and physics of Gromov--Witten theory and topological A-model  with target an algebraic variety $X$, and consider extensions involving pairs $(X,D)$ as above. We then present the first of our correspondences, relating the genus-0 log and local Gromov--Witten theories of the pair. An impasse is reached when trying to uplift the correspondence to higher genera, and a solution for this is proposed by invoking a relation to the all-genus open Gromov--Witten theory of an associated Aganagic--Vafa pair $(Y,L)$ in the form of a higher genus log-open correspondence. Special care, and a somewhat more extended treatment compared to the other parts of the manuscript, will be given to the process recasting maximal contact invariants of surfaces in the form of open Gromov--Witten counts inside Calabi--Yau threefolds in Section~\ref{sec:logopen}. This then mediates a connection with the Donaldson--Thomas theory of a quiver via an instance of the ``branes--quivers'' correspondence of \cite{Ekholm:2018eee,Kucharski:2017ogk,Panfil:2018faz}, as well as to other BPS invariants. Consequences for log Gromov--Witten theory, and generalizations to singular surfaces and non-nef divisors are finally discussed in Section~\ref{sec:non-nef}.

\section{Introduction}	

\subsection{Gromov--Witten theory}
\label{sec:gwintro}

\subsubsection{Gromov--Witten invariants: the geometry}

A typical way to tackle enumerative questions in algebraic geometry is to cast them as suitable intersection theory problems. In the case of counts of curves this proceeds by assigning, to a given algebraic variety $X$ (the ambient space of the counting problem), a suitable moduli space $\cM(X)$ parametrizing curves in $X$. The answer to the desired enumerative question can then be phrased as a suitable integration against the fundamental class of $\cM(X)$, with the integrand reflecting the incidence conditions relevant for the counting problem. The main trouble with this strategy however is that it is very rarely available: the sought-for moduli space $\cM(X)$ is almost invariably singular and non-compact, with different compactifications giving rise to different invariants. \\

The main type of compact moduli space we will look at arises from seeing curves in $X$ as \emph{parametrized curves} -- i.e. as maps from a source complex projective curve, modulo automorphisms of the domain. This leads to considering a moduli space
$\cM_{g,n}(X,d)$
parametrizing degree$-d$ maps from a smooth stable genus-$g$ pointed curves $(C; p_1, \dots p_n)$; here stability means that either $d\neq 0$, or $(C; p_1, \dots p_n)$ has zero-dimensional automorphism group\footnote{That is, $n \geq 3$ for $g=0$, and $n\geq 1$ for $g=1$.}. Its Kontsevich compactification $\overline{\cM}_{g,n}(X,d)$ consists of maps from possibly nodal sources, such that each contracted component is stable in the ordinary sense, counting nodes as marked points. This is a compact algebraic orbifold (proper Deligne--Mumford stack) of expected dimension 
\beq \mathrm{vdim} \overline{\cM}_{g,n}(X,d) = (\mathrm{dim} X-3)(1-g) - K_X \cdot d+n.
\label{eq:vdim}
\eeq
Although usually singular, reducible, and of impure dimension, the fact that $ \overline{\cM}_{g,n}(X,d)$ carries a perfect obstruction theory \cite{MR1437495} implies that it has a virtual fundamental class in the expected dimension $[ \overline{\cM}_{g,n}(X,d)]^{\rm vir} \in H_{2\mathrm{vdim} \overline{\cM}_{g,n}(X,d)}( \overline{\cM}_{g,n}(X,d))$; there are furthermore canonical evaluation morphisms 
\bea
\mathrm{ev} :  \overline{\cM}_{g,n}(X,d) & \longrightarrow & X^n \nn \\ 
\big[\phi : (C; p_1, \dots, p_n) \to X\big]
& \longrightarrow & (\phi(p_1), \dots, \phi(p_n)).
\eea
Given a collection of closed subvarieties $B_i$, this allows to define numbers
\begin{eqnarray}
n_{X,g,d}[B_1, \dots, B_n] &=& \hbox{``$\#$ degree-$d$, genus-$g$ curves in $X$ through $B_i$''} 
\nn \\
&:=& \int_{[ \overline{\cM}_{g,n}(X,d)]^{\rm vir}} \prod_{i=1}^n \mathrm{ev}_i^* [B_i]^\vee
\label{eq:gwdef}
\end{eqnarray}
where $[B_i]^\vee$ is Poincar\'e-dual to the homology class $[B_i]\in H_\bullet(X, \bbZ)$ of $B_i$. \\

The numbers $n_{X,g,d}[B_1, \dots, B_n]$ are the \emph{Gromov--Witten invariants of $X$}. The scare-quotes in  \eqref{eq:gwdef} are due to the possible existence of multiple cover contributions, usually preventing these invariants to be directly enumerative.

\begin{example}
Let's go back again to Question~{\bf Q3} in Section~\ref{sec:over}. Heuristically, the answer to it should be given by the Gromov--Witten count 
\beq
K_d \coloneqq n_{\bbP^2,0,d}[\stackrel{3d-1}{\overbrace{\mathrm{pt}, \dots, \mathrm{pt}}}]
\label{eq:kd}
\eeq 
and this is, in fact, one of the special circumstances where these invariants return the actual enumerative count \cite{MR1492534}.  Underlying the relatively classical look of the content of Question~{\bf Q3} is a range of surpisingly complex properties satisfied by the associated numbers. First of all, the geometry of boundary divisors on $\overline{\cM}_{0,n}(\bbP^2,d)$ imposes the existence of a non-linear recursion in the degree:
\beq 
K_d = \sum_{d_A+d_B=d}K_{d_A} K_{d_B} d_A^2 d_B \left(d_B
\binom{3d-4}{3d_A-2}-d_A \binom{3d-4}{3 d_A-1} \right) 
\label{eq:Nd}
\eeq
The resulting sequence (OEIS A013587) grows factorially, $K_d \sim (3d-1)! x_0^d$ for some $x_0 \in \bbR$. Despite efforts dating from the early days of topological field theory,  no analytic closed-form expression is currently available for either $x_0$ or the numbers $K_d$. The genus-0 Gromov--Witten potential of $\bbC\bbP^2$ is defined as the convergent power series
\beq
F_0^{\bbC\bbP^2}(t_1,t_2,t_3) = \frac{t_1^2 t_3}{2}+\frac{t_1 t_2^2}{2} + \sum_{d > 0} \frac{K_d}{(3d-1)!} \re^{d t_2} t_3^{3d-1}.
\label{eq:F0P2}
\eeq
The recursion \eqref{eq:Nd} is the reflection, at the level of Taylor coefficients, of the WDVV equations satisfied by $F_0$. 
Although no closed form expression is known for \eqref{eq:F0P2}, it is known that this is a special transcendental function, the non-linear WDVV recursion \eqref{eq:Nd} translating into a special case of Painlev\'e VI \cite{Dubrovin:1994hc}.
\vspace{.5cm}\end{example}
\subsubsection{Gromov--Witten invariants: the physics}
\label{sec:physics}
The Gromov--Witten counts have a physical interpretation as worldsheet instanton contributions to A-twisted topological correlators of a $\cN=(2,2)$ $\sigma$-model coupled with topological gravity \cite{Witten:1991zz}, having the smooth algebraic variety/K\"ahler manifold $X$ as its target. Denoting by $\phi^I$ and $g_{IJ}$, $I,J=1, \dots, 2d$ the local components of respectively a map $\phi :\bbC \to X$ and the K\"ahler metric in a real chart for $X$,  the $\sigma$-model is described by the $\cN=(2,2)$ supersymmetric action

\begin{equation}
S = 2 t \int_{\bbC}\rd z \rd \bar z\left(\frac{1}{2}g_{IJ}\partial \phi^{I}\bar{\partial} \phi^{J} + i\psi^{\bar{i}}_{-}D_{z}\psi^{i}_{-}g_{i\bar{i}} + i\psi^{\bar{i}}_{+}D_{\bar{z}}\psi^{i}_{+}g_{i\bar{i}} + R_{i\bar{i}j\bar{j}}\psi^{i}_{+}\psi^{\bar{i}}_{+}\psi^{j}_{-}\psi^{\bar{j}}_{-}\right) 
\end{equation}
where $\psi_{+/-}$ are left/right moving worldsheet fermions valued in the holomorphic/antiholomorphic tangent bundle of $X$, $D_{z/\bar{z}}$ are the covariant holomorphic/anti-holomorphic Dirac operators combining the spin connection on the worldsheet with the pull-back of the Levi--Civita connection on the target, and $R$ is the Riemann curvature tensor of the K\"ahler metric. The theory is invariant under supersymmetry transformations generated by four worldsheet supercharges living on the worldsheet, $Q^{\pm}$, and $\overline{Q}^{\pm}$, and has accordingly left and right classical $U(1)$ R-symmetries. In particular there is a non-anomalous vector current $U(1)_V=U(1)_L+U(1)_R$ which allows to define a topological twist of the theory: this redefines the Euclidean $\mathrm{SO}(2)$ rotation group on the worldsheet by the addition of the generator of the vector $R$-symmetry. Under the topologically twisted Euclidean rotation group, the supercharge $Q_A \coloneqq Q^+ + \overline Q^-$ has spin-zero and is therefore akin to a BRST operator; furthermore, the resulting action $S_A$ is $Q_A$-exact up to a topological term
\beq
S_A = -t \l(\int_\bbC \phi^*(\omega) + \{Q_A, V\}\r)
\eeq
where $\omega$ is the K\"ahler class, and $V=i \int_{\bbC} \rd z \rd\bar z g_{IJ} (\de_z \phi^I \de_{\bar z} \phi^J-\de_{\bar z} \phi^I \de_{z} \phi^J)$. The explicit $Q_A$ action on fields gives an isomorphism of graded differential modules between the BRST cohomology and the de Rham cohomology of $X$, the de Rham grading being identified with the vector $R$-symmetry charge. \\

The $Q_A$-exactness of the action has two main consequences:
\bit
\item The worldsheet theory, appropriately covariantized with respect to a background worldsheet metric, with a $Q_A$-invariant vacuum and once restricted to the cohomology of $Q_A$, is {\it topological}: the $Q_A$-exactness of the action implies the $Q_A$-exactness of the energy momentum tensor, implying that vacuum expectation values of $Q_A$-closed operators are constant on the worldsheet. 
\item For the same reason, the worldsheet theory is semi-classical: an infinitesimal variation in $t$ is a $Q_A$-exact operator insertion, which again vanishes in the $Q_A$-closed subsector of the worldsheet Hilbert space.
\eit
The last point implies that the worldsheet path integral heuristically localizes, with probability one, on on-shell/$Q_A$-invariant field configurations: in the scalar sector, this implies that $\phi:\bbC \to X$ is holomorphic. The $\sigma$-model can then be covariantized and coupled to worldsheet topological gravity on a closed oriented Riemann surface $C_g$, and its observables calculated as a string path integral modulo super-diffeomorphisms. In the $Q_A$-invariant sector and restricting to matter fields, i.e. for insertions corresponding to cohomology classes $\{[B_i]^\vee \in H^\bullet(X, \bbC)\}_{i=1}^n$ pulled back from the target manifold $X$, the corresponding observables decompose as a sum over worldsheet instantons (holomorphic maps, modulo worldsheet automorphisms) with discrete sectors labelled by the genus $g$ and the degree $d=\phi_*[C_g]\in H_2(X, \bbZ)$: this is the physical worldsheet realization of the Gromov--Witten invariants in \eqref{eq:gwdef}. \\

When the target $X$ is a Calabi--Yau threefold, a physical target space interpretation of the generating functions of Gromov--Witten invariants is also available as computing certain F-terms in a type IIA compactification on $\bbR^{1,3} \times X$. 
As the virtual dimension \eqref{eq:vdim} of the moduli space $\overline{\cM}_{g,0}(X,d)$ vanishes to all genera and degree, generating functions of GW invariants of $X$ can be computed as
\beq
\cF_{g}^X(t_i) \coloneqq  \sum_{d\in \mathrm{H}_2(X,\bbZ)} e^{-t (\omega, d)} n_{X,g,d}
\label{eq:fg}
\eeq
where fixing  $\{C_i\}_{i=1}^{h^{1,1}(X)}$ a set of generators of the effective cone of $X$ we denoted $t_i = t \int_{C_i} \omega = (\omega, d)$. Identifying $t_i$ with classical background field values for the vector multiplets of the type IIA effective $\cN=2$ supergravity on $\bbR^{1,3}$, the Gromov--Witten generating functions in \eqref{eq:fg} compute, for $g>0$, effective terms of the form
\beq
\int \rd^4x \rd^\theta \cW^{2g} \cF^X_g(t_i) = \int \rd^4 x \cF_g^X(t_i) R_+^2 F_+^{2g-2}
\eeq
where $R^2_+$ is a self-contraction of the self-dual part of the Riemann tensor, and $F_+=F+*F$ is the self-dual part of the graviphoton curvature. The genus 0 GW invariants of $X$ instead determine the prepotential
\beq
\int \rd^4 x (\de^2_{ij}\cF^X_0(t)) F_i^+ \wedge F_j^+  .
\eeq
where $F_i^+$ is the self-dual component of the field strength for the $U(1)$ gauge field in the $i^{\rm th}$ vector multiplet, $i=1, \dots, h^{1,1}(X)$.
\subsection{Looijenga pairs}
\label{sec:lp}
We would now like to raise our stakes a little, and consider enumerative invariants and associated topological string amplitudes that are sensitive not just to the geometry of a target algebraic variety, but also of a  subvariety in it of complex codimension~one. We start by giving the following definition.
\begin{definition}
A \emph{nef log~Calabi--Yau (CY)} pair is a pair $(X,D)$ with 
\bit 
\item $X$ a smooth complex projective variety;
\item  $D\in |-K_X|$ an effective anti-canonical divisor in $X$ with simple normal crossings singularities, admitting a decomposition $D=D_1 \cup \dots D_l$ with each $D_i$ irreducible, smooth, and nef (i.e. $C \cdot D_i \geq 0$ for all effective curves $C$ in $X$).
\eit
We will further say that $(X,D)$ is \emph{toric} is $X$ is a toric variety and $X \setminus D \simeq (\bbC^\star)^{\dim_\bbC X}$ is the big torus orbit.
\label{def:lp}
\end{definition}
\begin{definition}
A \emph{nef Looijenga pair} is a nef log~CY pair with $X$ a surface, $\dim_\bbC X=2$, and $D$ a singular divisor in $X$.
\end{definition}
\begin{remark}
Note that, since we require each irreducible component of $D$ to be smooth, and $D$ itself to be singular, our definition of a nef Looijenga pair requires $D$ to be reducible ($l>1$).
\end{remark}
\begin{example}
A few low-dimensional examples of nef log~CY pairs are as follows. 
\bit
\item $X=\bbC\bbP^1$, $D=\{0\}+\{\infty\}$ ($l=2$).
\item $X=\bbC\bbP^2$, $D=E$ with $E$ a smooth cubic ($l=1$). 
\item $X=\bbC\bbP^2$, $D=H \cup Q$ with $H$ a line, and $Q$ a conic ($l=2$)
\item $X=\bbC\bbP^2$, $D=$ the union of the coordinate axes ($l=3$)
\eit
The first and the last examples are toric ($D$ being the toric boundary); and the third and the fourth are Looijenga pairs since $D$ is singular. The second example is neither a Looijenga nor toric pair.
\label{ex:pairs}
\vspace{.5cm}\end{example}
By \cite[Propositions~2.2 and 2.3]{Bousseau:2020fus}, nefness of $D$ and smoothness of $X$ entail that there is a finite catalogue of eighteen smooth deformation families of nef Looijenga pairs. We will stick to $(X,D)$ being a Looijenga pair from now on, and consider two superficially different classes of enumerative invariants of a Looijenga pair $(X,D)$: the local and the (log) maximal contact invariants.
\section{Enumerative invariants}

\subsection{Local Gromov--Witten theory}
\label{sec:locgw}
Consider the vector bundle $\pi: E_{(X,D)} \to X$ on $X$, defined as the total space of the direct sum of the dual line bundles to the irreducible components $D_i$ of $D$:
\beq
E_{(X,D)} \coloneqq \mathrm{Tot}\big(\oplus_{i=1}^l \cO_X(-D_i) \big)
\eeq
$E_{(X,D)}$ is a smooth quasi-projective variety of dimension $l+2$, and since $D$ is anticanonical we have that $c_1(E_{(X,D)}) = \pi^*c_1(TX) + \pi^*c_1(\cO_X(-D)) = 0$, so that $E_{(X,D)}$ is a non-compact Calabi--Yau (CY) manifold.
\begin{figure}
\centering
 \includegraphics[scale=0.5]{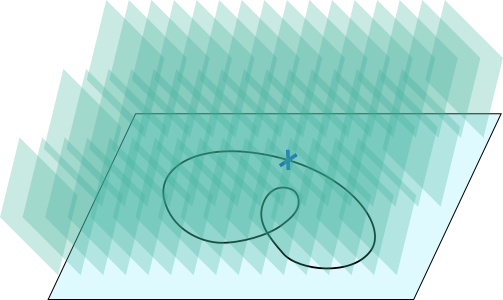}
\caption{A depiction of $E_{(X,D)}=\mathrm{Tot}(\oplus_{i=1}^l \cO_X(-D_i))$.}
\end{figure}
\begin{example}
For $(X,D)=(\bbC \bbP^2,H+Q)$, we have $E_{(\bbC \bbP^2,H+Q)} = \mathrm{Tot}(O_{\bbC\bbP^2}(-1)\oplus O_{\bbC\bbP^2}(-2))$. This is a non-compact CY fourfold.
\vspace{.5cm}\end{example}
The class of invariants we will be looking at, informally, should count the number of degree-$d$, genus $g=0$ curves in the total space $E_{(X,D)}$ satisfying a number of point conditions pulled back from $X$. Since $D_i$ is nef for all $i$, the image of a stable map intersecting $D_i$ generically will not be allowed to deform holomorphically off the zero section. Hence, as a scheme, the moduli space of maps to $E_{(X,D)}$ is just the moduli space of maps to $X$, and is therefore compact despite $E_{(X,D)}$ not being so. However the corresponding obstruction theories differ \cite{MR1437495}, the discrepancy being encoded into a canonical obstruction sheaf $\mathrm{Ob}_{X,D,d} \coloneqq R^1\pi_* f^* N_{X/E_{(X,D)}}$ on  $\overline{\cM}_{0,n}(X,d)$ \cite{Chiang:1999tz,MR2276766}, where
\beq
\xymatrix{ \mathcal{C}_{X,D} \ar[d]^\pi\ar^{f}[r]  &  X  \\
  \overline{\cM}_{0,n}(X,d) &}
\eeq
is the universal curve with its evaluation to $X$.  The virtual fundamental class for the local theory is defined as the intersection of the usual virtual fundamental class with the top Chern class of the obstruction bundle
\beq
[\overline{\cM}_{0,n}(E_{(X,D)},d)]^{\rm vir} = [\overline{\cM}_{0,n}(X,d)]^{\rm vir} \cap c_{\rm top}\big( \mathrm{Ob}_{X,D,d}\big).
\eeq
Since $\rank \mathrm{Ob}_{X,D,d}  = \dim H^1(\bbC\bbP^1, \oplus_i \phi^*\cO_X(-D_i))= -K_X \cdot d-l$, the virtual dimension for the local problem is $\mathrm{vdim}\overline{\cM}_{0,n}(E_{(X,D)},d)=n+l-1$. The corresponding \emph{local Gromov--Witten invariants} of $(X,D)$, virtually enumerating rational curves through $l-1$ points on the surface $X$, are defined as
\beq
N^{\rm loc}_{0,d}(X,D) \coloneqq \int_{[\cM_{0,l-1}(E_{(X,D)},d)]^{\rm vir}}\prod_{i=1}^{l-1}\mathrm{ev}_i^* [\mathrm{pt}].
\label{eq:locdef}
\eeq 
The invariants in \eqref{eq:locdef} have the following informal interpretation: let $\pi:\cY\to X$ be a projective CY $(2+l)$-fold containing a rigid surface $X$ with normal bundle $N_{X/\cY} \simeq E_{(X,D)}$, and let $d$ be the homology class of the image of a genus zero stable map to $\cY$ which is wholly contained in $X$. Then $N^{\rm loc}_{0,d}(X,D)\coloneqq\int_{[\cM_{0,l-1}(\cY,d)]^{\rm vir}}\prod_{i=1}^{l-1}\mathrm{ev}_i^* [\pi^*\mathrm{pt}_X]$ is the degree-$d$, genus zero genus GW invariant of $\cY$ with $l-1$ point conditions pulled back from $X$.

\subsection{Log Gromov--Witten theory}
We can also consider a different set of enumerative invariants attached to the pair $(X,D)$, where the counting occurs directly in $X$, but we use $D$ to impose ``boundary'' conditions for our map. We will be interested in virtually enumerating degree-$d$, rational curves in the surface $X$ passing through a suitable number of points, and having maximal tangency with each irreducible component $D_i$ of $D$:  note that for $(X,D)=(\bbC\bbP^2, H \cup Q)$ this is exactly the setup of Example~\ref{ex:p2lcintro}. 

An immediate problem one runs into is the inherent non-compactness of such a moduli space: the reason for this is that a sequence of stable maps with prescribed contact orders at $D_i$ may be allowed to splinter off some irreducible components that fall entirely into the divisor on the boundary of the moduli space, so that the contact condition does not make sense any longer in the limit. There are multiple dialects of Gromov--Witten theory that have been devised in order to keep track of the additional discrete data of the ramification of the divisors while at the same time sidestepping the issues with lack of compactness highlighted above, whether by considering stable maps into expanded degenerations of the target \cite{MR1938113, ranganathan2019logarithmic}, or  by considering some enhancement of the stable maps by extra combinatorial date using logarithmic geometry \cite{GS13,AbramChen14}. We will resort to the latter, viewing $X$ as a log scheme for the divisorial log structure induced by $D$: informally, we endow the datum of the stable map with discrete data tracking the tangency condition in terms of a homomorphism of lattice cones supported on the contact points. Referring the reader to \cite{GS13} for an extended survey of the construction, the resulting moduli space of log stable maps $\overline{\cM}^{\rm log}_{0,l-1}((X,D),d)$ is a proper log Deligne--Mumford stack, which furthermore \cite{GS13,AbramChen14} carries (under suitable minimality conditions) a perfect obstruction theory and a virtual fundamental class of expected dimension $n+l-1$ -- note that this is the same virtual dimension as for the local problem above. The desired count can then be defined as the log~Gromov--Witten invariant 
\beq
N^{\rm log}_{0,d}(X,D) \coloneqq \int_{[\overline{\cM}^{\rm log}_{0,l-1}((X,D),d)]^{\rm vir}}\prod_{i=1}^{l-1}\mathrm{ev}_i^* [\mathrm{pt}].
\label{eq:nlogdef}
\eeq
The log invariants \eqref{eq:nlogdef} are often enumerative -- and indeed coincide with the corresponding genuine count of curves with tangency conditions for Looijenga pairs since the interior $X\setminus D$ is a cluster variety \cite{Man19}. In particular, for $(X,D)=(\bbC\bbP^2, H \cup Q)$, they return the count \eqref{eq:ndintro} of rational plane curves maximally tangent at a line and a conic:
\beq
\mathfrak{n}_d =N^{\rm log}_{0,d}(\bbC\bbP^2,H\cup Q). 
\label{eq:nd=ndlog}
\eeq 

\begin{example}
It is interesting to contrast the calculation of the log and local invariants of $(X,D)$ with the ordinary GW invariants of $X$ -- see Table~\ref{tab:logloc}.
%
\begin{table}[!h]
\tbl{Log and local GW invariants of $(\bbC\bbP^2, H \cup Q)$ (left) and $(\bbC\bbP^2, H_1 \cup H_2 \cup H_3)$ (right), compared with the Kontsevich numbers $K_d$ in \eqref{eq:kd}.}{
\centering
\label{tab:logloc}
\begin{tabular}{c|c|c|c}
$d$ & $N_{0,d}^{\rm log}$ & $N_{0,d}^{\rm log}/N_d^{\rm loc}$ & {\color{gray} $K_d$}\\
\hline
\hline
 1 & 2 & $-2$ & {\color{gray}$1$}\\
\hline
 2 & 6 & 8 & {\color{gray}1}\\
\hline
 3 & 20 & $-18$ & {\color{gray}12} \\
\hline
 4 & 70 & 32  & {\color{gray}620} \\
\hline
 5 & 252 & $-50$ & {\color{gray}87304} \\
\hline
 6 & 924 & 72 & {\color{gray}26312976}\\
$\vdots$ & $\vdots$ & $\vdots$ & $\vdots$
\end{tabular}
~~~~
\begin{tabular}{c|c|c|c}
$d$ & $N_{0,d}^{\rm log}$ & $N_{0,d}^{\rm log}/N_{0,d}^{\rm loc}$ & {\color{gray} $K_{d}$}\\
\hline
\hline
 1 & 1 & $1$ & {\color{gray}$1$}\\
\hline
 2 & 4 & $-8$ & {\color{gray}1}\\
\hline
 3 & 9 & $27$ & {\color{gray}12} \\
\hline
 4 & 16 & $-64$  & {\color{gray}620} \\
\hline
 5 & 25 & $125$ & {\color{gray}87304} \\
\hline
 6 & 36 & $-216$ & {\color{gray}26312976}\\
$\vdots$ & $\vdots$ & $\vdots$ & $\vdots$
\end{tabular}
}
\end{table}
Even though the geometric setup is more complex here than in ordinary GW theory owing to the presence of the background divisor $D$, the numerology associated to the sequence of invariants is substantially (and surpisingly) simpler.
\bit
\item 
For $(X,D)=(\bbC\bbP^2, H \cup Q)$, the growth of the log invariants $N_{0,d}^{\rm log}$ is only exponential, $N_{0,d}^{\rm log}\sim 4^d$, instead of factorial.
\item Moreover, unlike the more mysterious sequence of absolute invariants $K_d$, they do seem to fit a recognisable pattern given by the OEIS sequence A000984 (the central binomial coefficients $\binom{2d}{d}$). This indicates that, unlike the ordinary GW invariants of $X$, the local and log invariants of $(X,D)$ might be amenable to an explicit, closed-form, non-recursive solution to all degrees.
\item Furthermore, even more recognisable is the growth of the ratio $N_{0,d}^{\rm log}/N_{0,d}^{\rm loc}$, which up to a sign factor is given by a quadratic law for this $l=2$ component case. 
\item Finally, the contrast with the ordinary GW theory of the plane is even starker for $(\bbC\bbP^2,L_1 \cup L_2 \cup L_3)$, where the log invariants are just $N_{0,d}^{\rm log}=d^2$, and the ratio with the local invariants is now given (up to a sign) by a cubic law for this $l=3$ component case.
\eit
\vspace{.5cm}\end{example}

These features are in fact by no means special to the example of $(\bbC\bbP^2, H \cup Q)$. In particular the power law behavior of the ratio of log to local invariants was found to be satisfied in a large variety of examples in \cite{vGGR}. This was inferred to a general conjecture, whose specialization to nef log CY surface pairs and point insertions is given by the following statement \cite{vGGR}.
\begin{conjecture}
For a nef log CY surface pair $(X,D)$, 
\beq 
N_{0,d}^{\rm log}(X,D) = \Bigg[ \prod_{i=1}^{l} (-1)^{d \cdot D_i+1} (d \cdot D_i)\Bigg] N_{0,d}^{\rm loc}(X,D).
\label{eq:vggr}
\eeq
\label{conj:vggr}
\end{conjecture}
The proposed equality in \eqref{eq:vggr} is a numerical version for point insertions of the \emph{log-local correspondence} proposed at the level of virtual classes in \cite{vGGR}. 

At face value, the very existence of the equality in \eqref{eq:vggr}, purporting an identity of local and maximal contact invariants up to a universal factor, is very unexpected. The two types of counts are {\it a priori} completely unrelated: the local count occurs in $2+l$ dimension, since as discussed at the end of \cref{sec:locgw} they encode the local contribution to the GW theory of a CY-$(2+l)$-fold of a of a rigid surface $X$ in it with normal bundle $\oplus_{i=1}^l \cO_X(-D_i)$; the log count instead is genuinely two-dimensional, and is further enriched with the datum of tangency conditions along the divisors. Furthermore, the local invariants are known to be rational numbers as they involve multi-covering contributions \cite{Chiang:1999tz,Klemm:2007in}, whilst the log invariants are integers and actually enumerative for Looijenga pairs \cite{Man19}. Still, evidence in favor of \eqref{eq:vggr} comes from the case $l=1$, established in \cite{vGGR} through a degeneration argument in log GW theory, and from an explicit solution of both sides of the equality for toric pairs in \cite{Bousseau:2019bwv}; and indeed, a stronger statement can in fact be made for Looijenga pairs \cite{Bousseau:2020fus}.
\begin{theorem}
Conjecture~\ref{conj:vggr} holds for nef Looijenga pairs. Moreover, both sides of the equality are closed-form solvable.
\label{thm:loglocal}
\end{theorem}
The statement of the Theorem claims simultaneously 1) a comparison result and 2) an explicit non-recursive solution for both types of invariants. A sketch of the ideology of the proof, broken down accordingly in these two parts, is as follows.

\subsubsection{Theorem~\ref{thm:loglocal}: the comparison}
\label{sec:comparison}
This part of the statement can be proved through a degeneration argument in log Gromov--Witten theory, imitating the analogous strategy adopted in \cite{vGGR} for the smooth case; we just give a sketch of the idea here, referring the reader to \cite{Bousseau:2020fus} for details, and to \cite{GarrelNotts} for a very readable survey.
The key idea is to degenerate the total space $E_{(X,D)}$ to a trivial bundle $X\times\bbA^l$ glued along $D_j\times\bbA^l$, $j=1,\dots, l$, to a rank-$l$ vector bundle over the projective bundle $\bbP(\cO_{D_j}\oplus \cO_{D_j}(-D_j))$, generalizing the idea of \cite{vGGR} for smooth divisors. The resulting family admits a log~smooth desingularization, to which the Abramovich--Chen--Gross--Siebert decomposition formula \cite{abramovich2017decomposition} can be applied: this expresses the local invariants $N^{\rm loc}_{0,d}(X,D)$ as a weighted sum of terms, indexed by tropical curves $h \colon \Gamma \to \Delta$,
where $\Delta$ is the dual intersection complex of the central fiber:
\beq N^{\rm loc}_{0,d}(X,D)
=\sum_{h \colon \Gamma \rightarrow \Delta} 
\frac{m_h}{|\mathrm{Aut}(h)|} 
N^{{\rm loc}, h}_{0,d}(X,D)\,.
\label{eq:nlocdec}
\eeq
The resulting equality would then follow from showing that the r.h.s. reproduces the expected relation \eqref{eq:vggr}. For $l=2$, it was shown in \cite{Bousseau:2020fus} that if $h$ is a non-maximal tangency type, the corresponding contribution can be shown to vanish, leaving out a single computable contribution from the maximal tangency term $h=h_{\rm max}$, for which 
\beq
N^{{\rm loc}, h_{\rm max}}_{0,d}(X,D)=\prod_{i=1}^{l}\frac{(-1)^{d\cdot D_i +1}}{(d \cdot D_i)^2} N^{{\rm log}}_{0,d}(X,D)
\eeq
and the multiplicity $m_{h_{\rm max}} = \prod_{i=1}^l d \cdot D_i$ yields exactly the expected proportionality factor in \eqref{eq:vggr}.

\subsubsection{Theorem~\ref{thm:loglocal}: the calculation}
\label{sec:calculation}

Two distinct calculational schemes are available for the local and log invariants. The local invariants are computed using the Coates--Givental theorem \cite{MR2276766}, which in genus zero expresses the local Gromov--Witten invariants of $(X,D)$ in terms of the ordinary genus zero descendent invariants of $X$ through an explicit hypergeometric modification of its $J$-function (see \cite{MR2510741} for general formulas, and Examples~\ref{ex:locgwHQ} and \ref{ex:locgwHHH} for two basic examples).
 Now all nef Looijenga pairs $(X,D)$ admit $\bbQ$-Gorenstein deformations to $(X',D')$ with $X'$ a smooth {\it toric} Fano surface \cite{Bousseau:2020fus}, for which $J_{E_{(X',D')}}(t,z)$ can be determined using Givental-style mirror theorems \cite{MR1653024}; furthermore, it turns out that the mirror map $z(t)$ is an explicit rational function of $\re^t$. By deformation invariance, the descendent invariants with a single point insertion can then be computed as
\beq
\Big[z^{1-l} \re^{t d}\Big]  J^0_{E_{(X',D')}} =  \int_{[\cM_{0,1}(E_{(X,D)},d)]^{\rm vir}}\mathrm{ev}^* [\mathrm{pt}] \psi_1^{l-2}.
\label{eq:nlocpsi}
\eeq
where $J^0$ is the identity component of the $J$-function. For $l>2$, the multi-point, non-descendent invariants $N_{0,d}^{\rm loc}(X,D)$ can be computed from \eqref{eq:nlocpsi} using a small-to-big quantum cohomology reconstruction theorem; see \cite{Bousseau:2019bwv,Bousseau:2020fus} for details, and Example~\ref{ex:locgwHHH} for an instance of the calculation.

\begin{example}
Let $(X,D)=(\bbC\bbP^2, H \cup Q)$; note that $X$ is already toric, and Givental mirror symmetry can be applied directly to it.
Consider the diagonal $\bbC^*$-action on the fibers of $E_{(\bbC\bbP^2,H \cup Q)}=\mathrm{Tot}(\cO_{\bbC\bbP^2}(-1)\oplus \cO_{\bbC\bbP^2}(-2))$: the mirror map in this case is trivial, and the corresponding equivariant $J$-function equates the equivariant $I$-function, which is given as
\beq
J_{E_{(\bbC\bbP^2,H \cup Q)}}(t,z)
= z \re^{t H/z} \sum_{d \in \bbZ_{\geq 0}} \re^{t d} \frac{\prod_{m=0}^{d-1} (-\lambda + H + m z) \prod_{m=0}^{2d-1} (-\lambda + 2H + m z)}{\prod_{m=1}^d (H+ m z)^3} 
\eeq
where $H=c_1(\cO_{\bbC\bbP^2}(1))$ is the class of a line. From \eqref{eq:nlocpsi} we get
\beq 
N_{0,d}^{\rm loc}(\bbC\bbP^2, H \cup Q) = [z^{-1} \re^{t d}]  J_{E_{(\bbC\bbP^2,H \cup Q)}}\Big|_{H =0} = \frac{(-1)^d}{2d^2} \binom{2d}{d},
\label{eq:locgwHQ}
\eeq
which recoves a direct calculation by virtual localization in \cite{Klemm:2007in}.
\label{ex:locgwHQ}
\vspace{.5cm}\end{example}

\begin{example}
Let $(X,D)=(\bbC\bbP^2, H \cup H \cup H)$. As for the case of the ordinary quantum cohomology of $\bbC\bbP^2$, it is not difficult to prove that 
\beq
\int_{[\overline{\cM}_{0,2}(E_{(X,D)},d)]^{\rm vir}}\mathrm{ev}_1^* [\mathrm{pt}] \mathrm{ev}_2^* [\mathrm{pt}] = \int_{[\overline{\cM}_{0,3}(E_{(X,D)},d)]^{\rm vir}}\mathrm{ev}_1^* [\mathrm{pt}]   \psi_1 \mathrm{ev}_2^* [H] \mathrm{ev}_3^* [H] \nn \\
\eeq
expressing the fact that the small quantum cohomology product $H^n \star H^m$  is equal to the cup product $H^{n+m}$ if $n+m<3$ (see \cite{Bousseau:2019bwv}). Applying twice the Divisor Axiom to the r.h.s.,  the 2-pointed local GW invariants of $(\bbC\bbP^2, H \cup H \cup H)$ are then related to the J-function,
\beq
J_{E_{(\bbC\bbP^2,H \cup H \cup H)}}(t,z)
= z \re^{t H/z} \sum_{d \in \bbZ_{\geq 0}} \re^{t d} \frac{\prod_{m=0}^{d-1} (-\lambda + H + m z)^3}{\prod_{m=1}^d (H+ m z)^3} 
\label{eq:JHHH}
\eeq
\beq 
N_{0,d}^{\rm loc}(\bbC\bbP^2, H \cup H \cup H) =  [z^{-1} \re^{t d}] \de_t^2 J_{E_{(\bbC\bbP^2,H \cup H \cup H)}}\Big|_{H =0} = d^2 [z^{-1} \re^{t d}]  J_{E_{(\bbC\bbP^2,H \cup H \cup H)}}\Big|_{H =0},
\eeq
from which we deduce that
\beq
N_{0,d}^{\rm loc}(\bbC\bbP^2, H \cup H \cup H) = d^2 \frac{(-1)^{d+1}}{d^3} = \frac{(-1)^{d+1}}{d}.
\label{eq:locgwHHH}
\eeq 
\label{ex:locgwHHH}
\vspace{.5cm}\end{example}

For the maximal tangency log invariants $N_{0,d}^{\rm log}(X,D)$, a systematic computational framework is provided by their associated scattering diagrams \cite{GPS,GHKlog,Gro11}. To a Looijenga pair $(X,D)$ one can (non-uniquely) construct pairs $(\widetilde{X},\widetilde{D})$ and $(\overline{X},\overline{D})$ fitting into a diagram \cite{GHKlog}
    \begin{equation}
    \label{eq:toricmodel}
        \begin{tikzcd}
            & (\widetilde{X}, \widetilde{D}) \arrow{ld}[swap]{\phi} \arrow{rd}{\pi} & \\
            (X,D) & & (\overline{X}, \overline{D}) 
        \end{tikzcd}
    \end{equation}
    where $\phi$ is a sequence of blow-ups at nodes of D and $\pi$ is a {\it toric model}, meaning that $(\overline{X}, \overline{D})$ is toric and $\pi$ is a sequence of blow-ups at distinct smooth points of $\overline{D}$. 

The log~GW invariants of $(X,D)$ can be related to the more easily computed invariants of $(\overline{X},\overline{D})$, as follows. First of all, by the log-birational invariance result of \cite{AW}, we have $N_{0,d}^{\rm log}(X,D)=N_{0,\tilde{d}}^{\rm log}(\widetilde{X}, \widetilde{D})$, where $\tilde d$ denotes the total transform. For a suitable divisor $F=A_1(\widetilde{X})$, the pair $(\widetilde{X}, \widetilde{D})$ can be degenerated to the normal cone of $F$ into a reducible pair with two components, one of which is the toric pair $(\overline{X}, \overline{D})$. The degeneration formula of \cite{abramovich2017decomposition} then expresses $N_{0,d}^{\rm log}(X,D)$ as a linear combination, with computable coefficients $c_\mathrm{m} \in \bbQ$, of the log GW invariants of $N_{0,d,\mathrm{m}}^{\rm log}(\overline{X}, \overline{D})$ with maximal contact $d \cdot D_i$ over
$\pi_*(D_i)$ and arbitrary ramification profile over $\pi_*(F)$ specified by a partition $\mathsf{m} \vdash d \cdot \pi_*(F)$: 
\beq
N_{0,d}^{\rm log}(X,D) = \sum_{\mathsf{m} \vdash (d \cdot \pi(F))} c_{\mathsf{m}} N_{0,d,\mathsf{m}}^{\rm log}(\overline{X}, \overline{D}).
\eeq 
The toric log~GW invariants on the r.h.s. can then be computed by correspondence theorems with tropical geometry in terms of a count of certain tropical curves in the fan of $\overline{X}$; see \cite{Mikh05,NiSi,ManRu,Man19}, and also \cite{GarrelNotts} for a nice survey.
  
It is helpful to consider the $l=2$ and $l>2$ cases separately. For the former, as discussed at the end of \cref{sec:comparison}, \cite[Theorem~5.1]{Bousseau:2020fus} provides a proof of the equality in \eqref{eq:vggr}; and \cite[Proposition~3.2 and Theorem~3.3]{Bousseau:2020fus} gives a general closed formula for the local invariants via local mirror symmetry techniques, thanks to the fact that the mirror map is closed-form invertible for all $(X,D)$. As the number of components $l$ increases, the degeneration to the singular fiber becomes increasignly singular, and the comparison argument accordingly more involved; however the separate calculation of the local and log invariants simplifies considerably in this more degenerate setting, meaning that a closed formula can be found for both sides of \eqref{eq:vggr}, from which the sought-for equality can be deduced.

\begin{example}
Consider again the case $(X,D) = (\bbC\bbP^2, H \cup Q)$.  Since $l=2$, the comparison result of \cite[Theorem~5.1]{Bousseau:2020fus} for 2-component Looijenga pairs gives
\beq
N_{0,d}^{\rm log}(\bbC\bbP^2, H \cup Q) = (-1)^{d} 2 d^2  N_{0,d}^{\rm loc}(\bbC\bbP^2, H \cup Q),
\eeq
and using \eqref{eq:locgwHQ} we obtain that the maximal tangency log invariants,  returning the curve count $\mathfrak{n}_d$ in \eqref{eq:ndintro}, are given by
\beq
\mathfrak{n}_d = N_{0,d}^{\rm log}(\bbC\bbP^2, H \cup Q) = \binom{2d}{d}.
\label{eq:nlogp2hq}
\eeq 
\vspace{.5cm}\end{example}

\begin{figure}[h]
\begin{center}
\begin{tikzpicture}[smooth, scale=1.2]
\draw[step=1cm,gray,very thin] (-2.5,-2.5) grid (2.5,2.5);
\draw (0,0) to (-2.5,0);
\draw (0,-2.5) to (0,0);
\draw (0,0) to (2.5,2.5);
\draw[thick] (1,-1) to (-2.5,-1);
\draw[thick] (1,-2.5) to (1,-1);
\draw[thick] (1,-1) to (2.5,0.5);
\node at (-0.25,-1.8) {$T_2$};
\node at (1.4,1.7) {$T_3$};
\node at (-1.7,0.2) {$T_1$};
\node at (-1,-1) {$\bullet$};
\node at (1,-2) {$\bullet$};
\node at (-0.75,-0.8) {$P_1$};
\node at (1.25,-1.8) {$P_2$};
\node at (1.2,-2.3) {$d$};
\node at (-2.3,-0.8) {$d$};
\node at (2.4,0.2) {$d$};
\node at (0.9,-0.8) {$d^2$};
\end{tikzpicture}
\end{center}
\caption{The unique genus 0, degree $d$, maximally tangent tropical curve in $\mathrm{Fan}(\bbC\bbP^2)$ passing through two general points $P_1$ and $P_2$. It carries weight $d^2$.}
\label{fig:p2hhh} 
\end{figure}
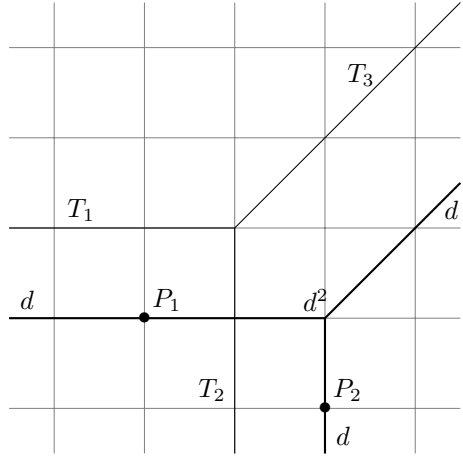

\begin{example}
Consider now the 3-component Looijenga pair $(X,D) = (\bbC\bbP^2, H \cup H \cup H)$. In this case, given that $l>2$, it is more expedient to compute the log invariants directly, and deduce the correspondence with the local invariants by direct comparison with \eqref{eq:locgwHHH}. Note that since $X$ is toric, and $D$ is the toric boundary, the pair $(X,D)$ coincides with its toric model $(\overline{X}, \overline{D})$, and the log invariants are then directly computed as a tropical count. In particular,
\beq
N_{0,d}^{\rm log}(\bbC\bbP^2, H \cup H \cup H) = \sum_\Gamma  \mathrm{Mult}(\Gamma) 
\eeq
where the sum on the r.h.s. runs over genus 0, degree $d$  maximally tangent tropical curves $\Gamma$ through 2 points in $\mathrm{Fan}(\bbC\bbP^2)$. These are trivalent trees in the fan of $\bbC\bbP^2$ whose edges $e$ have rational slope and carry an integer weight $w(e)$, such that there exist exactly three unbounded edges decorated with weight $d$ parallel to the rays of the fan, and the compact edges satisfy the balancing condition $\sum_{e \ni v} w(e) u_{(v,e)} =0$ where $u_{(v,e)}$ is the primitive outgoing vector parallel to $e$. The multiplicity $\mathrm{Mult}(\Gamma)$ is defined as
\beq 
\mathrm{Mult}(\Gamma) \coloneqq \prod_{v \in \Gamma} w(e) w(e') | \det(u_{(v,e)} u_{(v,e')}) |
\eeq 
with $e \neq e' \ni v$; this is well-defined by the balancing condition. In the case of $(\bbC\bbP^2, H \cup H \cup H)$ there is only one such tropical curve $\Gamma$, depicted in \cref{fig:p2hhh}, for which $\mathrm{Mult}(\Gamma) = d^2$. Hence,
\beq
N_{0,d}^{\rm log}(\bbC\bbP^2, H \cup H \cup H) = d^2 = (-1)^{d+1} d^3 N_{0,d}^{\rm loc}(\bbC\bbP^2, H \cup H \cup H) 
\eeq 
as expected.
\vspace{.5cm}\end{example}

\subsection{A higher genus puzzle}
The above discussion was entirely confined to genus zero; a natural question is thus how much of it could be transferred to the setting of higher genus GW theory. From the virtual dimension formula \eqref{eq:vdim}, for $\dim X=2$ the expected dimension of the moduli space of genus-$g$ stable maps to $X$ is $g$ higher than the genus-0 virtual dimension, and the same occurs for the virtual dimension of the log moduli space $\overline{M_{g,n}^{\rm log}}(X,D; d)$:
\beq
\mathrm{vdim}\overline{M_{g,n}^{\rm log}}(X,D; d) = 
\mathrm{vdim} \overline{M_{0,n}^{\rm log}}(X,D; d)+g=g+n+l-1.
\eeq
Usually, a zero-dimensional virtual count in higher genus is defined by compensating the increase in virtual dimension by additional incidence conditions pulled back from the target. Alternatively, in this case one could cap the virtual class with the top (degree-$g$) Chern class of the Hodge bundle, $\lambda_g \coloneqq c_{\rm top} (R^1 \pi_* \cO_{\mathcal{C}_{g,n}})$ where $\pi : \mathcal{C}_{g,n} \to \overline{M^{\rm log}}_{g,n}(X,D; d)$ is the universal curve, as already suggested in \cite{GPS}. The corresponding higher genus log GW invariants are then defined as
\beq
N_{g,d}^{\rm log}(X,D)  \coloneqq  \int_{[\overline{\cM}_{g,l-1}^{\rm log} (X,D,d)]} \prod_{i=1}^{l-1} \mathrm{ev}_i^* [\mathrm{pt}] (-1)^g \lambda_g.
\label{eq:nlog_g}
\eeq
It will be convenient to package these into an all-genus generating function, defined as
\beq
\bbN_{d}^{\rm log}(X,D) \coloneqq \bigg(\frac{\hbar}{2\sin(\hbar/2)}\bigg)^{l-2} \sum_{g \geq 0} (\hbar)^{2g} N_{g,d}^{\rm log}(X,D)\,.
\label{eq:nlog_gf}
\eeq
The enumerative significance of the higher genus invariants \eqref{eq:nlog_gf} with $\lambda_g$ insertions was elucidated by Bousseau in \cite{bousseau2018quantum}, building upon \cite{MR3904449,filippini2015block}: as in genus zero, these invariants can be related to a weighted count of tropical curves in the corresponding toric model, with the tropical multiplicity being replaced with $[\mathrm{Mult}_\Gamma]_q$, where $q=e^{\ri \hbar}$ and $[n]_q \coloneqq \frac{q^{n/2}-q^{-n/2}}{q^{1/2}-q^{-1/2}}$ is the symmetric $q$-analogue of $n \in \bbZ_{\geq 0}$.	 In particular, the same calculations for the log invariants in $g=0$ can be leveraged to give all-genus results for the generating function in \eqref{eq:nlog_gf}. 
\begin{example}
Let $(X, D)=(\bbC\bbP^2, H \cup Q)$. In this case, the $q$-deformed version of the tropical calculation of the log-invariants gives the $q$-analogue of \eqref{eq:nlogp2hq},
\beq
\bbN_{d}^{\rm log}(\bbC\bbP^2,H \cup Q) = \qbinom{2d}{d}_q 
\label{eq:p2hqg}
\eeq
where $\qbinom{m}{n}_q$ is the $q$-binomial coefficient $[m]_q!/([n]_q! [m-n]_q!)$, and $[n]_q \coloneqq \prod_{i=1}^n [i]_q$ is the symmetric $q$-factorial.
\vspace{.5cm}\end{example}

An obvious question then is how the log-local correspondence of \cref{thm:loglocal} can be extended to all genera. The question is actually moot in itself: the expected dimension of the virtual class for stable maps into a CY $(2+l)$-fold is $(l-1)(1-g)$, which is negative for $g>1$ and no marked points, and in particular there are no non-vanishing local GW invariants for higher genus. \\

\begin{figure}[t]
\centering \includegraphics[scale=0.75]{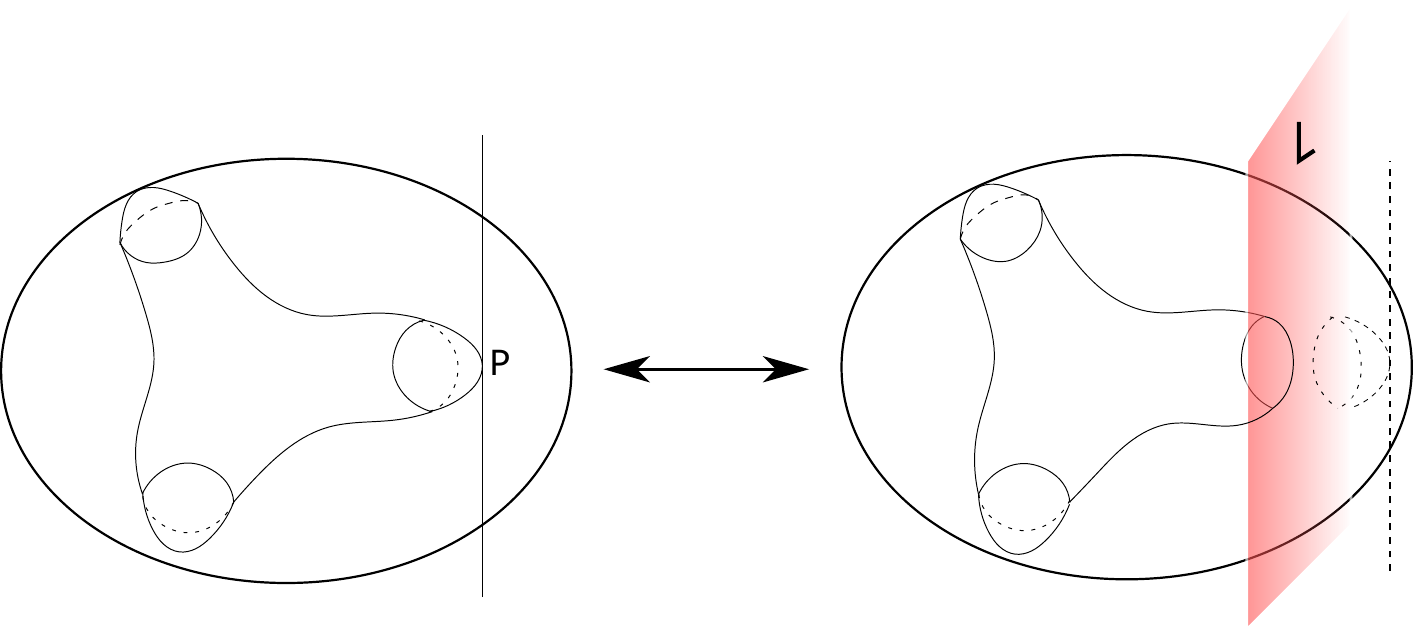}
\caption{Tangency conditions on divisors versus open conditions on Lagrangians.}
\label{fig:logopen}
\end{figure}

\subsection{Log vs open invariants and QFT engineering}
\label{sec:logopen}

A solution to the resulting impasse was proposed in \cite{Bousseau:2020fus} by proposing that the CY$(2+l)$-fold local GW invariants $N_{0,d}^{\rm log}(X,D)$ coincide with some open GW invariants virtually counting open stable maps into a local CY3-fold, with boundaries lying on special Lagrangian submanifolds $L_1 \cup .. \cup L_{l-1}$. There is both a topological and a physical rationale for this. Geometrically, and from the log perspective, a stable map with a contact condition at a point on a divisor $D$ can be naturally seen as a limiting version of an open stable map with an open boundary condition on a special Lagrangian $L$ with a homologically non-trivial $S^1$ which bounds a holomorphic disk emerging from $D$: as the Lagrangian is pushed against the divisor, boundaries close up into punctures, and the winding number around the circle gives the ramification around the contact point (see \cref{fig:logopen}). Physically, and from the local perspective, the proposed identification of log and open invariants is a higher dimensional version of the geometric engineering of quantum field theories of \cite{Katz:1996fh,MR2183121,Kachru:1995fv}: genus zero {\it closed} topological string amplitudes in a CY {\it fourfold} are known to compute certain superpotential F-terms in an effective compactification of type IIA to two dimensions, with four supercharges \cite{Gukov:1999ya}, in the same vein as the four-dimensional protected terms discussed in Section~\ref{sec:physics}. At the same time, the same type of F-terms can be engineered as a topological {\it disk} amplitude on a {\it threefold} by wrapping D4-branes around special Lagrangians in a CY3-fold \cite{Ooguri:1999bv}. It was realized in \cite{Mayr:2001xk} (see \cite{Liu:2021eeb,Liu:2022swc} for a recent in-depth mathematical study) that for some  local geometries the {\it same} effective theories can sometimes be engineered in {\it either} way: therefore, whenever this physical equivalence occurs we obtain a geometrical identity between the corresponding (closed 4-dimensional and open 3-dimensional) enumerative invariants!  Indeed, for $l\geq 2$ and under relatively mild conditions \cite{Bousseau:2020fus}, it is possible to show that 
\ben
\item for the local geometries associated to a nef Looijenga pair $(X,D_1 \cup \dots \cup D_l)$, one can systematically associate a corresponding open string CY3 geometry $(Y, L= L_1 \cup \dots L_{l-1})$, where $Y$ is a quasi-projective CY3 variety and $L_i \subset Y$, $i=1, \dots, l-1$ are special Lagrangian submanifolds;
\item the $(l-1)$-holed open GW counts on $Y$ with boundaries ending on $L_1, \dots, L_{l-1}$ are equal to closed string counts with point insertions on the local Calabi--Yau $(2+l)$-fold $E_{(X,D)}$;
\item unlike the closed string invariants, the open ones admit a zero-dimensional virtual fundamental class at all genera, corresponding to gravitational $F$-terms involving higher powers of the Weyl multiplet and the gaugino superfield. We propose that those are exactly what ``refines'' the local invariants of $E_{(X,D)}$, and the correspondence of Theorem~\ref{thm:loglocal} with the log invariants, at higher genus. 
\een
\subsubsection{Genus zero: from local to open invariants}
\begin{figure}[!h]
\centering
\includegraphics[scale=0.8]{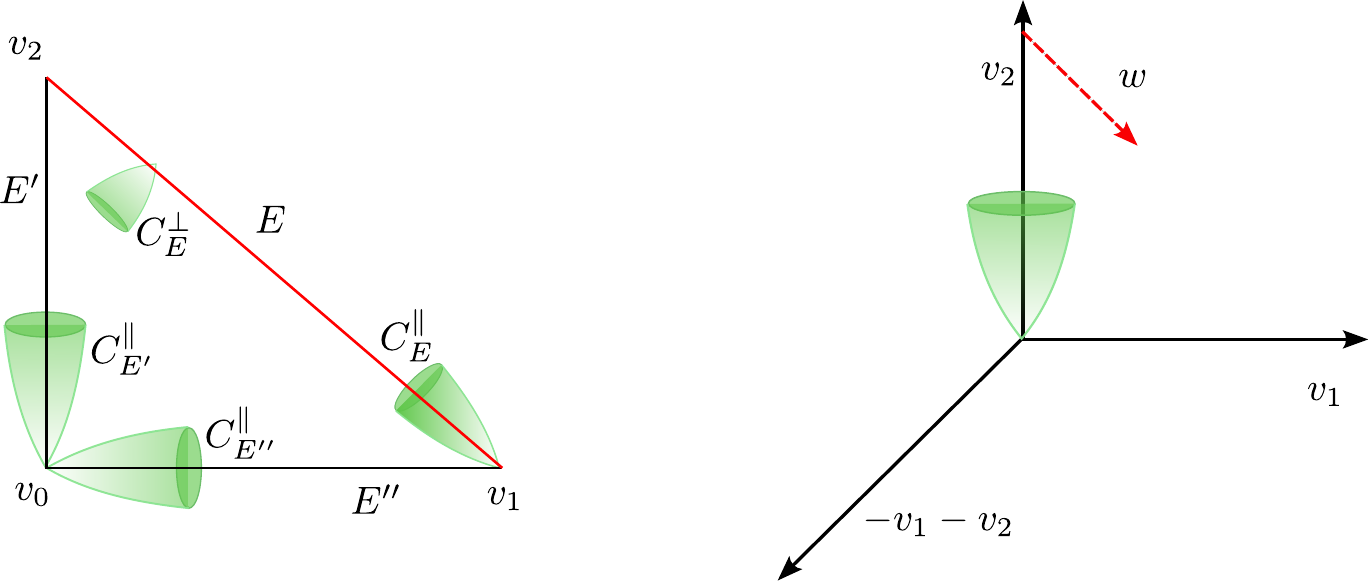}
\caption{The moment polytope of $\bbC\bbP^2$ (left) with a depiction of the generators $C_{E}^\|$ and $C_E^\perp$ (resp. $C_{E'}^\|$ and $C_{E''}^\|$) of the homology of the torus fiber $\ell$; and the corresponding toric diagram of $\bbC^3$ (right) with a toric brane on the edge $E'$ at framing $f=1$. Note that the framing vector $w=v_1-f v_2= v_1-v_2$ is parallel to $E$.}
\label{fig:log2openpoly}
\end{figure}

So how is the special Lagrangian pair $(Y,L)$ constructed from $(X,D)$? The idea here is to employ a mixture of the heuristic expetations linking (log) invariants with fixed ramification to
\ben[(i)]
\item  on one hand, open invariants with winding number equal to the contact order at the divisor;
\item on the other, local invariants of the surface geometry twisted by the total space of the canonical bundle.
\een
So suppose, for simplicity and to simplify notation in the following, that $X$ is toric, $l=2$, and that $D_1$ is a prime toric divisor: for example take our running example of the projective plane $X=\bbC\bbP^2$ with $D_1=H$, $D_2=Q$ a line and a quadric. What we would like to do is to trade the maximal contact conditions along $D_i$, $i=1,2$ with
\ben[(i)]
\item  for $i=1$: an open condition on a special Lagrangian near $D_1$.
\item  for $i=2$: a twist by $\cO(-D_2)$, as in \cite{vGGR};
\een
Now, since $X$ is a compact symplectic toric manifold, it is a Lagrangian torus fibration $\mu : X \to (\mathfrak{u}(1)^{\oplus 2})^* \simeq \bbR^2$ over a convex, bounded, and (up to scaling of the symplectic form) integral reflexive moment polygon $\Gamma$, given by the convex hull of lattice vectors $\{v_i \in \bbZ^2\}_{i=0}^{b_2(X)+1}$: the fibers over the codimension-$m$ stratum of the moment polygon are $2-m$ dimensional tori. Likewise, given that $D\in |-K_X|$, the total space $Y \coloneqq \mathrm{Tot}(\cO(-D_2)|_{X \setminus D_1})$ is a quasiprojective toric CY3, given by a $\bbT^2 \times \bbR$ (Harvey--Lawson) fibration over an $\bbR$-bundle over the polarization of the moment polytope of $X\setminus D_1$; in physics parlance, this is the ``pq-web'' or ``toric diagram'' of $Y$ \cite{Aganagic:2003db}, see Figure~\ref{fig:log2openpoly}. \\

Now suppose $\{v_1, v_2\}$ is an integral basis of $\bbZ^2$, and let $\{v_1^*, v_2^*\}$ be its dual basis for the Lie algebra of Hamiltonian vector fields associated to the fibration. Then the
one parameter group generated by $\exp t (a_1 v_1^* + a_2 v_2^*)$, $a_i \in \bbZ$, acts trivially on the fibers lying above the edges of $\Gamma$ with slope $(a_2, -a_1)$. Let's further denote $\ell \simeq \bbT^2 \subset X $ the torus fiber over an interior point $p \in \mathrm{Int}(\Gamma)$ near the edge $E \coloneqq \mu(D_1)$ of the moment polytope of $X$ corresponding to the toric divisor $D_1$, and let $L$ be the corresponding fiber of the Harvey--Lawson fibration of $Y$: note that $L$ is a trivial $\bbR$-bundle over $\ell$. The first homology of $\ell$ (and hence $L$) can be presented as $\bbZ [C^\|_E] \oplus \bbZ [C^\perp_E]$: here $C^\|_E$ is the equator of $D_1 \simeq \bbC\bbP^1$, and $C^\perp_E=\exp t v$ is the circle fiber generated by the Lie algebra element associated to the primitive outward pointing vector normal to the edge $E_i$: in particular, $C^\perp_E$ is the boundary of a disk intersecting $D_1$ at a point. 
\begin{example}
Let's cast the above definitions in the context of $(X,D)=(\bbC\bbP^2, H \cup Q)$, where $\bbC\bbP^2$ is equipped with the canonical symplectic form $\omega \coloneqq -2 \omega_{\rm FS}$, where $\omega_{\rm FS}$ is the Fubini--Study form. Then the moment map associated to the $\bbT^2$-action $[z_0 : z_1 : z_2] \to [z_0 : \re^{\theta_1}z_1 : \re^{\theta_2}z_2]$ is 
\beq
\mu[z_0 : z_1 : z_2] = \frac{1}{|z_0|^2+|z_1|^2+|z_2|^2} \big(|z_1|^2, |z_2|^2).
\eeq
The moment polytope $\Gamma$ in this case is the convex hull of $v_1=(1,0)$, $v_2=(0,1)$ and $v_0=(0,0)$ (see Figure~\ref{fig:log2openpoly}). The divisor given by the pre-images under the moment map of  the edge connecting $v_i$ and $v_j$ corresponds to the toric divisor $z_k=0$, $k \neq i,j$: in particular any edge represents the homology class of the line $D_1=H \in \mathrm{H}_2(\bbC\bbP^2,\bbZ)$. To see
what $\ell$ and its homology generators look like is in this case, take for definiteness $E$ to be the diagonal edge connecting $v_1$ and $v_2$. We have that $C^\|_E$ and $C ^\perp_E$ are, respectively, the orbit associated to $v_1^*-v_2^*$ and $-v_1^*-v_2^*$, i.e.
\beq
C^\|_E : \quad [z_0 : \re^{\theta} z_1 : \re^{-\theta} z_2] , \qquad C^\perp_E : \qquad [z_0 : \re^{\theta} z_1 : \re^{\theta} z_2] = [\re^{-\theta} z_0 : z_1 : z_2] 
\eeq
where $|z_i|\neq 0$ and $|z_0| \ll 1$ (i.e. $\ell$ is ``near'' the divisor). \\

The corresponding toric Calabi--Yau threefold is 
\beq
Y=\mathrm{Tot}(\cO_{X}(-D_2)|_{X \setminus D_1}) = \mathrm{Tot}(\cO_{\bbC\bbP^2}(-2)|_{\bbC\bbP^2 \setminus H}) \simeq \mathrm{Tot}(\cO_{\bbC^2}) = \bbC^3.
\eeq
The corresponding web diagram is obtained from the moment polytope of $\bbC\bbP^2$ by removing\footnote{Or rather by pushing it to infinity, corresponding to the infinite rescaling of the symplectic form that realizes $\bbC^2$ as a decompactification of $\bbC\bbP^2$.} the edge $E$, giving the moment polygon of $\bbC^2$, and by completing the origin into a balanced trivalent vertex with an edge pointing in the direction $(-1,-1)$.   
\label{ex:fromp2toc3}  
\vspace{.5cm}\end{example}
The heuristic physics expectation of \cite{Mayr:2001xk} can then be phrased as
\beq 
N_{0,d}^{\rm loc}(X,D) =O_{0, \iota(d)}(Y,L)
\label{eq:locopen}
\eeq
where the r.h.s. is a Gromov--Witten count of disks in a relative 2-homology class $\iota(d)$, described as follows. Let $i:X \setminus D_1 \hookrightarrow X$ be the open inclusion map. Consider the embedding of lattices 
%
%
%
\bea
\iota : {\mathrm H}_2(X, \bbZ) & \longrightarrow &  {\mathrm H}_2(X \setminus D_1, \bbZ) \bigoplus {\mathrm H}_1(\ell,\bbZ) \nn \\
d & \longrightarrow & (i^*(d); (d \cdot D_1) [C^\perp_E])\,,
\label{eq:maphom}
\eea
%
Then \eqref{eq:locopen} identifies the $4$-fold Gromov--Witten invariant of $E_{(X,D)}$ in class $d$ with a virtual count of disks in $Y$ in class $i^*(d)$ and with boundaries wrapping $L$ with winding number $0$ along $[C^\|_E]$ and $(d\cdot D_1)$ along $C^\perp_E$. Although the definition of open Gromov--Witten counts is notoriously daunting, we can take advantage here of the fact that $Y$ is toric, and consider a limit where the Lagrangian $L$ is deformed to a singular Harvey--Lawson fiber  with topology $\bbR^2 \times S^1$ (i.e. an \emph{Aganagic--Vafa brane} \cite{Aganagic:2000gs}). To this end, let $E'$, $E''$ be edges of the moment polygon $\Gamma$ incident to $E$, and let $f\in \bbZ$ be such that 
\beq 
[C_E^\perp]=[C_{E'}^\|]+f [C_{E''}^\|] \in H_1(L, \bbZ).
\eeq 
By standard toric arguments, note that $f$ equates the self-intersection number $\mu^{-1}(E)^2=D_1^2 = \deg N_{D_1/X}$. Consider now a degeneration of $\ell$ whereby its image on the moment polytope hits $E'$, so that $L \simeq \bbR^2 \times S^1$ is an Aganagic--Vafa brane, where the $S^1$ is homotopic to $C_{E'}^\|$, the equator of $\mu^{-1}(E)$.  Since the corresponding toric Lagrangian $L$ in $Y$ is the zero-locus of an anti-holomorphic involution, the moduli space of genus zero, 1-holed stable maps to $(Y, L)$ in absolute homology class $\beta \in H_2(Y, \bbZ) \simeq H_2(X\setminus D_1, \bbZ)$ and boundary class $\gamma [C_E']^\| \in {\rm H}_1(L, \bbZ) \simeq \bbZ$ admits a $\bbC^*$-equivariant virtual class in virtual dimension zero, and the corresponding invariants can be defined as
\beq
O_{0, (\beta,\gamma)}(Y,L) = \int_{[\cM_{0,\beta,\gamma}(Y,L)]^{\rm vir}} 1.
\label{eq:toricopengw}
\eeq
Two comments are in order:
\bit
\item first of all, in this toric limit, the map \eqref{eq:maphom} turns into a lattice isomorphism ${\mathrm H}_2(X, \bbZ) \simeq {\mathrm H}_2(Y, L,\bbZ) \simeq {\mathrm H}_2(Y,\bbZ) \oplus {\mathrm H}_1(L,\bbZ) \simeq {\mathrm H}_2(X \setminus D_1,\bbZ) \oplus {\mathrm H}_1(C_{E'}^\|,\bbZ) $ with $\iota(d)=(\beta, \gamma)$ and $\beta=i^*(d)$, $\gamma=d\cdot E$; the second in the string of isomorphisms here comes from the splitting of the relative homology sequence since ${\mathrm H}_2(L, \bbZ)=0$ for a toric brane, and the third by homotopy invariance upon retraction to the base; 
\item secondly, the attentive reader will notice at this point that, naively, \eqref{eq:toricopengw} reconstructs the disk invariants in the r.h.s. of \eqref{eq:locopen} only for $[C_E^\perp]=[C_{E'}^\|]$, i.e. when $f=0$. However the invariants are known to be affected by an integer {\it framing} ambiguity in the choice of $\bbC^*$-action in the localization, which in fact precisely reflects the framing relation between the class of the bounding holomorphic disk, $[C_E^\perp]$, and $[C_{E'}^\|]$! 
\eit
We will henceforth denote $L^{[f]}$ for the datum of an Aganagic--Vafa brane with a choice of framing $f$. and $O_{0, (\beta,\gamma)}(Y,L^{[f]})$ the corresponding disk Gromov--Witten invariant.
\beq 
N_{0,d}^{\rm loc}(X,D) =O_{0, \iota(d)}(Y,L^{[D_1^2]})
\label{eq:locopen2}
\eeq
\begin{example}
In the setting of \cref{ex:fromp2toc3}, we have $D_1^2=H^2=1$, so the toric Lagrangian pair associated to $(X,D)=(\bbC\bbP^2, H \cup Q)$ is $(Y,L)=(\bbC^3, L)$ where $L\simeq  \bbR^2 \times S^1$ is an Aganagic--Vafa brane at framing one. The isomorphism $\iota : \mathrm{H}_2(\bbC\bbP^2, \bbZ) \to  \mathrm{H}_2(\bbC^3, L, \bbZ) \simeq \mathrm{H}_1(L, \bbZ) \simeq \bbZ$ in this case simply sends $H \to [S^1]$.
\label{ex:fromp2toc3op}
\vspace{.5cm}\end{example}

The construction generalizes with only minor modifications to higher $l$, by replacing tangency conditions along $D_i$ with special Lagrangian open conditions near it -- we refer the reader to \cite{Bousseau:2020fus} for a more diffuse discussion. The general expectation, for any $l>1$, is that there exist framed toric Lagrangians $L=\sqcup_{i<l} L_i^{[f_i]}$ in a toric Calabi--Yau threefold $Y \simeq \mathrm{Tot}(\cO(-D_l)|_{X\setminus \cup_{i<l} D_i})$ and a canonical identification of ${\rm H}_2(X,\bbZ) \stackrel{\iota}{\simeq} {\rm H}_2(Y, \bbZ) \oplus_i {\rm H}_1(L_i^{[f_i]}, \bbZ)$  such that the following correpondence holds:
\begin{conjecture}
For $(X,D)$ and $(Y,L)$ as above, we have
\beq 
N_{0,d}^{\rm loc}(X,D_1 +\dots + D_l) =O_{0, \iota(d)}(Y,L_1^{[f_1]} \sqcup \dots \sqcup L_{l-1}^{[f_{l-1}]}).
\label{eq:locopengen}
\eeq
\label{conj:locopen}
\end{conjecture}

\begin{figure}[h]
\centering
\includegraphics[scale=0.8]{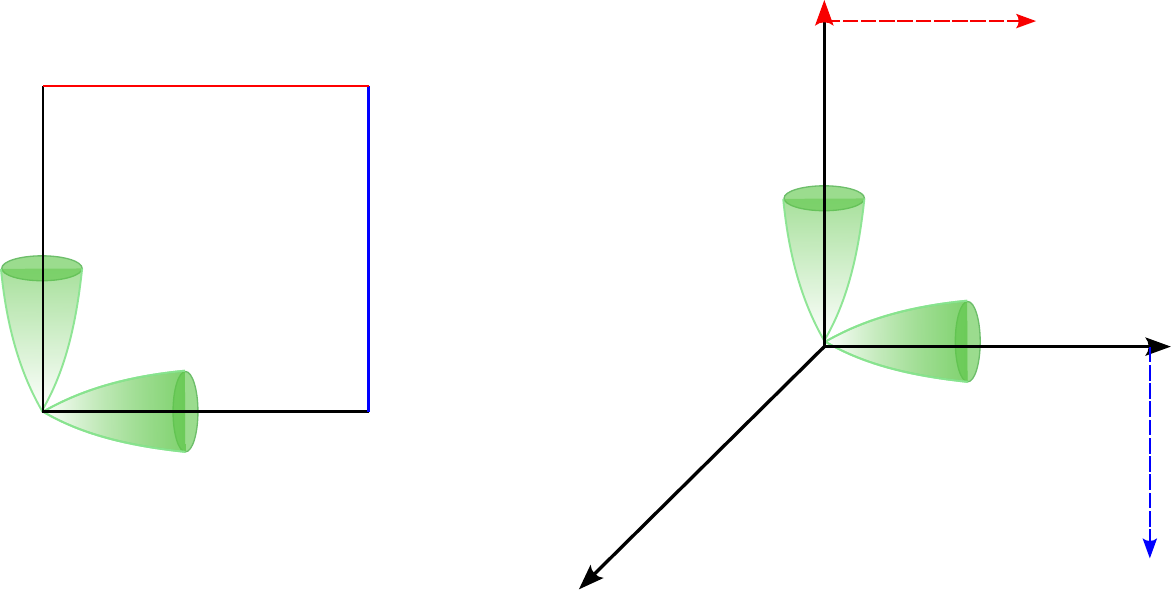}
\caption{The moment polytope of $\bbC\bbP^1\times \bbC\bbP^1$ (left), with the edges corresponding to $D_1$ and $D_2$ depicted in blue and red; and the corresponding toric diagram of $\bbC^3$ (right) with two toric branes corresponding to contact conditions along $D_1$ and $D_2$. Note that the framing vectors run parallel to the edges $\mu(D_1)$ and $\mu(D_2)$ that have been deleted from the polytope.}
\label{fig:p1xp1ann}
\end{figure}

\begin{example}
As an example with $l=3$, take $X=\bbC\bbP^1 \times \bbC\bbP^1$, and $D=D_1 \cup D_2 \cup D_3$ with $D_1=H_1$, $D_2= H_2$ ($H_i$ being the class of the $i^{\rm th}$ $\bbC\bbP^1$-factor), and $D_3$ a smooth member of the linear system generated by the diagonal $H_1+H_2$. Then we have that $Y=\mathrm{Tot} \cO(-D_3)|_{X \setminus \{D_1, D_2\}} =\mathrm{Tot} \cO(-1,-1)|_{\bbC\bbP^1 \times \bbC\bbP^1  \setminus \{H_1, H_2\}} \simeq \bbC^3$, and $L_i$ are Aganagic--Vafa Lagrangians on the edges $v_1=(0,1)$ and $v_2=(1,0)$ of the vertex with framing $0$ and $-1$ respectively -- see Figure~\ref{fig:p1xp1ann}. Denoting $[S^1_{(i)}]$ the generator of ${\rm H}_1(L_i, \bbZ)$, absolute homology classes of $X$ and relative homology classes of $(\bbC^3, L_1 \sqcup L_2)$ are identified as
\bea 
{\rm H}_2(\bbC\bbP^1 \times \bbC\bbP^1,\bbZ) & \stackrel{\iota}{\longrightarrow}  &
{\rm H}_1(L_1^{[0]} ,\bbZ) \oplus {\rm H}_1(L_2^{[-1]} ,\bbZ) \nn \\
d_1 H_1 + d_2 H_2 & \stackrel{\iota}{\longrightarrow} & d_1 [S^1_{(1)}]+d_2 [S^1_{(2)}]
\eea
and so \eqref{eq:locopengen} becomes
\beq 
N_{0,d_1 H_1 +d_2 H_2}^{\rm loc}(\bbC\bbP^1 \times \bbC\bbP^1) =O_{0,  d_1 [S^1_{(1)}]+d_2 [S^1_{(2)}]}(Y,L_1^{[0]},L_2^{[0]}).
\label{eq:locopenann}
\eeq
\vspace{.5cm}\end{example}

\begin{remark}
Central to our construction was the fact that the surface $X$ itself, as well as the divisors $D_i$ with $i<l$, are toric. It turns out that there is always a smooth deformation whose central fiber has these properties, and we can then use deformation invariance of Gromov--Witten invariants to specialize to this toric setting. In doing so, however, attention must be paid to the fact that the nefness of the divisor $D_l$ we utilize for the twisting in the construction of the Calabi--Yau threefold geometry $\mathrm{Tot} \l(\cO(-D_l)|_{X \setminus \{D_i\}_{i<l}}\r)$ is preserved under deformation to the toric model. There are in fact a handful of special cases in higher Picard number that don't satisfy this property, and which we won't consider further in this survey: the reader may find an extensive discussion in \cite{Bousseau:2020fus}.
\end{remark}

\subsubsection{Higher genus: from open to log invariants}

The relation \eqref{eq:locopengen} generalizes Mayr's open/closed string duality\footnote{This is very different from a duality in the usual sense of the gauge/string correspondence: there is no large $N$ in \cite{Mayr:2001xk}, the open theory being a $U(1)$ theory on a fixed worldsheet (a disk).} in \cite{Mayr:2001xk} to higher dimension and arbitrary framing, by relating the closed A-model on a class of CY-$(l+2)$-fold local surfaces to the open topological A-model with Dirichlet boundary conditions on $l-1$ Lagrangians in a toric Calabi--Yau threefold: in that light, the local surface GW invariants of $(X,D)$ are just a disguised form of genus zero {\it open} GW invariants on a CY {\it threefold}. One immediate advantage of the identification with the invariants of the associated threefold open string geometry is that, unlike $N_0^{\rm loc}(X,D)$, the theory of open stable maps to a toric special Lagrangian pair $(Y,L=\sqcup_{i<l} L_i^{[f_i]})$ has a moduli space carrying a virtual fundamental class of dimension zero {\it at all genera}, out of which we can define
\beq
O_{g, (\beta,\gamma)}(Y,L) \coloneqq \int_{[\cM_{g,\beta,\gamma}(Y,L)]^{\rm vir}} 1,
\label{eq:open_g}
\eeq
and as in \eqref{eq:nlog_gf} we can form an all-genus generating function
\beq
\bbO_{(\beta,\gamma)}(Y,L) \coloneqq \l(\frac{\hbar}{2 \sin(\hbar/2)}\r)^{l-3}\sum_{g \geq 0} \hbar^{2g} O_{g, (\beta,\gamma)}(Y,L),
\label{eq:open_gf}
\eeq
refining simultaneously the genus zero open and the local invariants in \eqref{eq:locopengen} to higher genera. \\

So how can this be used to refine the log-local correpondence to a higher genus log-open correspondence? Note that in genus zero, by \eqref{eq:locopengen} and Theorem~\ref{thm:loglocal}, it would follow that
\beq
N_{0,d}^{\rm log}(X,D_1 +\dots + D_l) = \l(\prod_{i=1}^l (-1)^{d \cdot D_i +1} d\cdot D_i\r) O_{0, \iota(d)}(Y,L_1^{[f_1]} \sqcup \dots \sqcup L_{l-1}^{[f_{l-1}]}).
\label{eq:logopen0}
\eeq
In the r.h.s., the factors $(-1)^{d \cdot D_i +1} d\cdot D_i$ can be heuristically interpreted as follows: for $i<l$, this arises from trading point conditions in $X$ by point conditions on $D_i$, with the latter corresponding to fixing a special Lagrangian condition near it: the sign factor is an intrinsic ambiguity in the open theory associated to the framing change $f_i \to -1-f_i$;
for $i=l$, this is just the predicted relative factor of the log-local correspondence in the irreducible case, arising from the twist by $\cO(-D_l)|_{X\setminus \cup_{i<l} D_i}$. While the former is genus-independent, the latter is corrected in higher genus as described by the higher genus log-local theorem in the irreducible case \cite{Bousseau:2020ckw}. The main result of \cite{Bousseau:2020ckw} consists of a beautiful (and intricated) relation between invariants of smooth {\it projective} Fano surfaces relative to a smooth anticanonical curve and the local invariants of the surface, expressed in terms of the invariants of the elliptic curve. Since $X\setminus \cup_{i<l} D_i$ is quasi-projective, a heuristic extrapolation of the arguments of \cite{Bousseau:2020ckw} indicate that the corresponding relation between (equivariant, under a Calabi--Yau torus action) relative and local invariants dramatically simplifies, and is expressed in terms of the higher genus $\lambda_g$-invariants of the {\it point}: at the level of generating functions, this is simply obtained by replacing $d \cdot D_l$ by its $q$-deformation \cite{Bousseau:2020fus}. Piecing everything together, this leads to the following
\begin{conjecture}
Let $(X,D)$ be a nef Looijenga pair admitting a deformation to a pair $(X',D')$ with $X'$ toric, $D'_i$ prime toric divisors for $i<l$, and $D'_l$ nef, and denote $(Y,L)$ the corresponding Aganagic--Vafa pair. Then,
\bea
\bbN_{d}^{\rm log}(X,D_1 +\dots + D_l) &=& \l(\prod_{i<l} (-1)^{d \cdot D_i +1} d\cdot D_i\r)  (-1)^{d\cdot D_l+1} [d \cdot D_l]_q\nn \\ 
& & \bbO_{\iota(d)}(Y,L_1^{[f_1]} \sqcup \dots \sqcup L_{l-1}^{[f_{l-1}]})\,,
\label{eq:logopeng}
\eea
where $q=e^{\ri \hbar}$.
\label{conj:logopeng}
\end{conjecture}
In the discussion getting us to Conjecture~\ref{conj:logopeng} we've been mainly guided by expectations (and slight generalizations) of the physics equivalence between local and open invariants, and some heuristics of the possible relation between open conditions on Lagrangians versus contact conditions on divisors. It turns out that our fantasy can be vindicated \cite{Bousseau:2020fus}, with a full higher genus analogue of Theorem~\ref{thm:loglocal}.
\begin{theorem}
Conjecture~\ref{conj:logopeng} holds, and moreover, both sides of the equality are closed-form solvable.
\label{thm:logopeng}
\end{theorem}
The proof follows from an explicit comparison of the calculation of $\bbN_{d}^{\rm log}$ and $\bbO_{\iota(d)}$ using, respectively, the $q$-deformed tropical vertex formalism \cite{bousseau2018quantum,MR3904449} and the topological vertex \cite{Aganagic:2003db,Li:2004uf}, carried out in \cite{Bousseau:2020fus, Bousseau:2020ryp, Brini:2022htq}.  By taking the genus zero limit, $q\to 1$, and using Theorem~\ref{thm:loglocal}, we immediately obtain that the generalized version of Mayr's duality holds.

\begin{corollary}
Let $(X,D)$ be a nef Looijenga pair as in Conjecture~\ref{conj:logopeng}. Then Conjecture~\ref{conj:locopen} holds,  and moreover, both sides of the equality are closed-form solvable.
\label{cor:locopen}
\end{corollary}

\begin{example}
Consider once again our pet example of $(X,D)=(\bbC\bbP^2, H \cup Q)$. From Examples~\ref{ex:fromp2toc3} and \ref{ex:fromp2toc3op}, the corresponding open string geometry $(Y,L)$ is given by $\bbC^3$ with a toric brane at framing 1. The generating function of all-genus, $1$-holed Gromov--Witten invariants at winding $d$ is given from the topological vertex \cite{Aganagic:2003db} as  
\beq
\bbO_{d}(Y,L) = \frac{1}{d}\sum_{R} \chi_R((d)) q^{\kappa(R)/2}(-1)^{d} s_R(q^{\rho}) 
\label{eq:openc3-1}
\eeq
where $R$ is an irreducible representation of the symmetric group $S_d$ labelled by a Young diagram with $d$-boxes, $s_R(q^\rho)$ is the corresponding Schur function in the principally stable specialization $s_R(x_i=q^{-i+1/2})$, $\kappa(R)$ is its second Casimir invariant,  and $\chi_R(c)$ denotes the character of the conjugacy class $[c]$ of $S_d$ in the representation $R$. Since $[c]=(d)$ is labelled by a full permutation cycle, the Murnaghan--Nakayama rule gives $\chi_{R}( (d) ) =(-1)^s$ if the Young diagram of $R$ is a hook diagram with $d$ boxes and $s+1$ rows, and zero otherwise. Using that (see e.g. \cite{MR325407})
%
$$s_{(d-s,1^s)}(q^\rho) = \frac{q^{\frac{1}{2}\l(\binom{d}{2}-d s\r)}}{[d]_q [d-s-1]_q! [s]_q!} \,,$$
%
we get that
\bea
\bbO_{d}(Y,L)
&=& \frac{(-1)^d}{d [d]_q}
 \sum_{s=0}^{d-1} (-1)^s  q^{\frac{3}{2} \binom{d}{2}}  \qbinom{d-1}{s}_q (-q^d)^{s} q^{-ds/2}   \nn \\
&=& \frac{(-1)^d}{d [2 d]_q } \qbinom{2d}{d}_q
\eea
where in the last line we have used the Cauchy binomial theorem. Comparing with \eqref{eq:p2hqg} returns exactly the expected relation in \eqref{eq:logopeng}.
\label{ex:p2hqvert}
\vspace{.5cm}\end{example}

\section{Applications}

Theorems~\ref{thm:loglocal} and \ref{thm:logopeng} have a host of non-trivial applications for the enumerative geometry of $(X,D)$. We explore some of these below.

\subsection{Logarithmic invariants from the topological vertex}
\label{sec:vertex}
\begin{figure}
\centering
\includegraphics[scale=1.5]{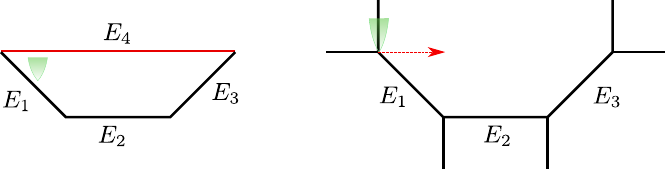}
\caption{The moment polygon (left) and the toric diagram of the associated open string geometry (right) for $X=\mathrm{dP}_3$, $D_1=H-E_1$, $D_2=2H-E_2-E_3$. The edge $E_4$ corresponding to $\mu(D_1)$ is highlighted in red; note that the framing vector, corresponding to $f=-1$, runs parallel to the deleted edge.}
\label{fig:dp3}
\end{figure}

The first application of Theorem~\ref{thm:logopeng} is the proof itself: as stated above, and as evidenced in Example~\ref{ex:p2hqvert}, the methods employed in \cite{Bousseau:2020fus, Bousseau:2020ryp, Brini:2022htq} rely on a direct {\it manu militari} calculation of both sides of the equality \eqref{eq:logopeng}, which give explicit formulas for the log and the open invariants separately. In some cases these formulas agree on the nose: this occurs when either $l>2$, or $l=2$ and $D_i^2>0$. But in general the two expressions are superficially very different, and the proof of \eqref{eq:logopeng} turns into a combinatorial problem in its own right.
\begin{example}
Let $\mathrm{dP}_3\coloneqq \mathrm{Bl}_{3 {\rm pts}} \bbC\bbP^2$ be the blow-up of the plane at three points. We write $H$ for the total transform of the line, and $E_i$, $i=1,2,3$ for the exceptional divisors. The anticanonical class has a decomposition $D=D_1+D_2$ with $D_1=H-E_1$ and $D_2=2H-E_2-E_3$, both having smooth effective representatives. By blowing up non-generically, we get the toric Fano surface with moment polytope $\mu : \mathrm{dP}_3 \to \bbR^2$ depicted in Figure~\ref{fig:dp3}: in particular the torus fibration of the top edge in the polytope is in the class $D_1$, which in this case is a prime toric divisor with $D_1^2=0$. Writing $d=d_0 (H-E_1-E_2-E_3)+ \sum d_i E_i$ for the class of an effective curve in $\mathrm{dP}_3$, the generating function of all-genus logarithmic invariants can be computed as the following intricate-looking multi-variate $q$-hypergeometric sum \cite{Brini:2022htq}
	\begin{equation}
			\begin{split}
				& \bbN^{\rm log}_d\big(\mathrm{dP}_3, D_1+D_2\big) = 
				  \quad \sum_{\substack{\forall (i,n)\in\{1,2,3,4\}\times\bbZ_{>0}: \, k_{i,n}\geq 0 \\ d_0 = \sum_{n\geq 1}\sum_{i=1}^4  (n+\delta_{i,1})  k_{i,n} \\ d_1 = \sum_{n\geq 1} \sum_{i=1}^4 k_{i,n} \\ d_0 - d_2 = \sum_{n\geq 1} (k_{1,n}+k_{4,n})  \\ d_0 - d_3 = \sum_{n\geq 1} (k_{1,n}+k_{3,n})}} ~\prod_{n\geq 1} \prod_{i=1}^{2} c_{i,n}(q) d_{i,n}(q)\,,
			\end{split}
			\label{eq:NlogdP302scat}
		\end{equation}
with 
\bea
c_{i,n} &=& \qbinom{d_2+d_3 - \sum_{m\geq1} \big(2m (k_{1,n+m} + k_{2,n+m}) + (2m-1) (k_{3,n+m} + k_{4,n+m})\big)}{k_{i,n}}_q\,, \nn \\
d_{i,n} &=& \qbinom{d_2+d_3 - \sum_{m\geq0} \big((2m+1) (k_{1,n+m} + k_{2,n+m}) + 2m (k_{3,n+m} + k_{4,n+m})\big)}{k_{2+i,n}}_q\,. \nn \\
\eea
On the other hand, the corresponding open string geometry $(Y,L)$ obtained by deleting the divisor $D_1$ and replacing the contact condition with a Lagrangian one is obtained on the r.h.s. of Figure~\ref{fig:dp3}, with a toric brane attached to an outer vertex at framing $-1$. A straightforward topological vertex calculation leads to 
\bea
\bbO_{\iota(d)}\l(Y,L\r) &=&
 \frac{(-1)^{d_1+d_2+d_3} [d_1]_q}{d_1 [d_0]_q [d_1+d_2+d_3-d_0]_q}\qbinom{d_3}{d_0-d_1}_q \nn \\
& \times & 
\qbinom{d_3}{d_0-d_2}_q
\qbinom{d_0}{d_3}_q
\qbinom{d_1+d_2+d_3-d_0}{d_3}_q,
\label{eq:opendp302}
\eea
where, in terms of the generators $[\mu^{-1}(E_i)]\in \mathrm{H}_2(Y, \bbZ)$, $i=1,2,3$, and $[S^1] \in \mathrm{H}_1(L,\bbZ)$ we have
\bea
\iota[H-E_1-E_2-E_3]=[C_3]-[C_1]\,, & \quad & \iota[E_1]=[C_1+C_2]\,, \nn \\
\quad \iota [E_2]=[C_1]\,, & \quad & \iota[E_3] =[S^1]+[C_1]-[C_3].
\eea
Then \eqref{eq:logopeng} turns into a new conjectural $q$-hypergeometric summation formula
\bea
\bbN^{\rm log}_d\big(\mathrm{dP}_3, D_1+D_2\big) &=& \frac{[d_1]_q [d_2+d_3]_q}{[d_0]_q [d_1+d_2+d_3-d_0]_q}\qbinom{d_3}{d_0-d_1}_q 
\qbinom{d_3}{d_0-d_2}_q \nn \\ & &
\qbinom{d_0}{d_3}_q
\qbinom{d_1+d_2+d_3-d_0}{d_3}_q\,.
\eea
An inductive proof {\it ex post}, based on the knowledge of the open invariants, was given in \cite{Brini:2022htq}, adapting arguments due to Krattenthaler in \cite{CKpriv}.
\vspace{.5cm}\end{example}

In general, the vertex calculation provides a closed form resummation of higher genus log GW generating functions for all nef Looijenga pairs under the conditions of Conjecture~\ref{conj:logopeng}.

\subsection{Open/closed BPS invariants, quivers, and integrality}
\label{sec:bpsdt}
A second application concerns the integral structure underlying the local invariants of $(X,D)$. Recall that $E_{(X,D)}$ is a (non-compact) Calabi--Yau $(2+l)$-fold. For these, a higher dimensional version of the genus zero Gopakumar--Vafa invariants of CY3 can be {\it defined} by the divisor sum 
\beq
\mathrm{KP}_{d}(X,D) = \sum_{k | d} \frac{\mu(k)}{k^{4-l}} N^{\rm loc}_{0,d/k}(X,D).
\label{eq:kp}
\eeq
mimicking the Aspinwall--Morrison multiple covering formula of the three-dimensional case. In \eqref{eq:kp}, $\mu(k)$ is the M\"obius function
\beq
\mu(k)=\l\{ \bary{rl}
 1   & \quad k \hbox{ is square-free and with an even number of prime factors,}\\
 -1   & \quad k \hbox{ is square-free and with an odd number of prime factors,}\\
 0  & \quad k \hbox{ has repeated prime factors.}
\eary
\r.
\eeq
When $l=2$, these were conjectured to be related to a count of BPS states, and are therefore integers, in work of Klemm and Pandharipande \cite{Klemm:2007in}. This integrality statement was generalized to higher dimensional Calabi--Yau varieties by Ionel and Parker, and proved in the compact case using symplectic methods, in \cite{MR3739228}. \\

\begin{figure}[h]
\centering
\begin{tikzpicture}[scale=2.4]
\node at (-3.6,0) {$\mathbb{C}^d$};
\node at (-2.2,0) {$\scriptstyle{\alpha\in\rm{End}(\mathbb{C}^d)\ni \beta}$};
\draw [fill] (-3.4,0) circle [radius=0.04];
\draw (-2.6,0) circle [x radius=0.8, y radius=0.6];
\draw (-2.6,-0.6) -- (-2.5,-0.5);
\draw (-2.6,-0.6) -- (-2.5,-0.7);
\draw (-3,0) circle [x radius=0.4, y radius=0.3];
\draw (-3,-0.3) -- (-2.9,-0.2);
\draw (-3,-0.3) -- (-2.9,-0.4);
\end{tikzpicture}
\caption{The quiver associated to $(X,D)=(\bbC\bbP^2, H\cup Q)$.}
\label{fig:quiverp2hq}
\end{figure}
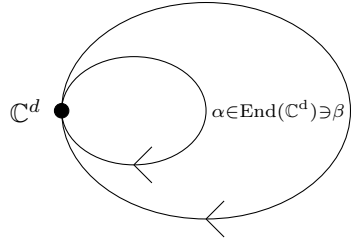

In our non-compact setup we can give an algebro-geometric proof of the integrality of Klemm--Pandharipande invariants as a direct corollary of Corollary~\ref{cor:locopen}, as follows. From the pioneering work of Ooguri and Vafa \cite{Ooguri:1999bv} and Labastida--Mari\~no--Vafa \cite{Labastida:2000yw}, open GW invariants are also conjectured to have an underlying integral structure, in terms of a count of open BPS bound states of M2-branes ending on M5-branes that wrap the framed toric Lagrangians $L=\cup_i L_i^{[f_i]}$. The open version multi-covering formula for $(l-1)$-holed amplitudes on a CY3 has structurally the same form of \eqref{eq:kp} on a CY-$(2+l)$:
\beq
\mathrm{LMOV}_{(\beta, \gamma)}(Y,L) = \sum_{k | (\beta, \gamma)} \frac{\mu(k)}{k^{4-l}} O_{0,(\beta/k, \gamma/k)}(Y,L).
\label{eq:lmov}
\eeq
By \eqref{eq:locopengen}, for a nef Looijenga pair, the associated local and open invariants coincide upon identifying $(\beta, \gamma)=\iota(d)$; and by \eqref{eq:kp} and \eqref{eq:lmov}, the associated BPS invariants are defined through the {\it same} multi-covering formula. Therefore we immediately find that
\beq
\mathrm{KP}_{d}(X,D)  = \mathrm{LMOV}_{\iota(d)}(Y,L)\,.
\label{eq:kplmov}
\eeq
On top of establishing a new link between open/closed BPS states living in different dimensions, \eqref{eq:kplmov} has also an immediate practical consequence as there is a lot more that is known about LMOV invariants than we know about KP invariants. For example, for $l=2$ the work of \cite{Kucharski:2017ogk,Ekholm:2018eee, Panfil:2018faz} associates to $(Y,L)$ a symmetric quiver $\mathsf{Q}(Y,L)$, which physically encapsulates the datum of a three-dimensional $\cN=2$ Abelian Chern--Simons-matter theory, which describes the dynamics on the worldvolume of an M5 brane wrapping the conormal bundle to the homologically non-trivial circle in $L$, and whose vortex partition function of this theory gives the generating series of numerical Donaldson--Thomas invariants of $\mathsf{Q}(Y,L)$. In particular, under the 3d-3d correspondence,  the open BPS invariants of \eqref{eq:lmov} coincide (up to sign) with the Donaldson--Thomas invariants of the corresponding quiver:
\begin{theorem}
Let $(X,D_1+D_2)$ be 2-component a nef Looijenga pairs satisfying the assumptions of Conjecture~\ref{conj:logopeng}, and let $(Y,L)$ be the associated Aganagic--Vafa special Lagrangian pair. Then there exists a symmetric quiver and an embedding of $\Delta : \mathrm{H}_2(Y,L, \bbZ) \hookrightarrow \bbZ[\mathsf{Q}(Y,L)_0] $ into the free abelian group generated by the vertices of the quiver such that
\beq
|\mathrm{LMOV}_{(\beta, \gamma)}(Y,L)| = \mathrm{DT}_{\Delta(\beta,\gamma)}\l(\mathsf{Q}(Y,L)\r)
\label{eq:lmovdt}
\eeq
where $\mathrm{DT}_{D}\l(\mathsf{Q}(Y,L)\r)$ denotes the numerical Donaldson--Thomas invariant of the quiver for the dimension vector $D\in \bbN[\mathsf{Q}(Y,L)_0]$.
\end{theorem}
The assignment of a quiver to $(Y,L)$ is non-unique, but for $L$ a toric brane there exist canonical minimal choices with number of vertices equal to the topological Euler characteristic of $Y$, and for which the identification of relative homology degrees with dimension vectors of the quiver is an actual isomorphism.
\begin{example}
Let $(X,D)=(\bbC\bbP^2, H \cup Q)$, so that $(Y,L)$ is given by $\bbC^3$ with a toric brane at framing one. In this case, since $\chi(Y)=1$ the minimal quiver $\mathsf{Q}(Y,L)$ has only one vertex, and the integral framing equal to one translates into the fact that $\mathsf{Q}(Y,L)$ is the 2-loop quiver \cite{Panfil:2018faz} -- see Figure~\ref{fig:quiverp2hq}. From \cite{MR2889742}, these invariants are equal to
\beq
\{\mathrm{DT}_d\l(\mathsf{Q}(Y,L)\r)\}_{d \in \bbZ_{>0}} = \{ 1, 1, 1, 2, 5, 13, 35, 100, 300, 925, 2915, 9386, \dots\}
\eeq
reproducing the (absolute value of the) Klemm--Pandharipande invariants in \cite[Section~3.2]{Klemm:2007in}.
\vspace{.5cm}\end{example}
For toric branes, the equality (up to sign) between LMOV invariants and quiver DT invariants can be rigorously established by a combined use of the topological vertex and CoHA methods to match the corresponding (higher genus vs motivic) $q$-hypergeometric series in the $q \to 1$ limit \cite{Panfil:2018faz}. The integrality of the DT invariants follows from a theorem of \cite{MR2956038}: this implies immediately, via \eqref{eq:kplmov} and \eqref{eq:lmovdt}, that the LMOV invariants and especially the KP invariants of the local geometry of the associated Looijenga pair are integral, proving the conjecture of \cite{Klemm:2007in} for local surfaces. At the same, not only are the $ N^{\rm log}_{0,d}(X,D)$ integers, since they are enumerative, but the open BPS/DT integrality gives a finer integrality statement for the corresponding log invariants: 
\begin{theorem}
We have that
\beq
\mathrm{KP}_d(X,D) = \sum_{k | d} \frac{\mu(k)}{k^{4-2l}} \prod_{i\leq l} \frac{(-1)^{d/k \cdot D_i+1}}{d \cdot D_i} N^{\rm log}_{0,d/k}(X,D) \in \bbZ.
\eeq
\end{theorem}
Furthermore, the existence of a non-trivial higher genus theory of the logarithmic and open invariants allows to make a refined statement in that setup. The prediction of \cite{Labastida:2000yw} is that the BPS generating function
\bea
 \mathbb{LMOV}_{(\beta,\gamma_1, \dots, \gamma_{l-1})}(Y,L) & \coloneqq & 
\prod_{i<l} \frac{\gamma_i}{[\gamma_i]_q} \sum_{k | (\beta, \gamma_1, \dots, \gamma_{l-1})} \frac{\mu(k)}{k} \mathbb{O}_{(\beta, \gamma_1, \dots, \gamma_{l-1})/k}(Y,L)(q^k)\,, \nn \\
\label{eq:lmovg}
\eea
is an integral Laurent polynomial in $q$. This can be proved directly (including for $l>1$) for all nef Looijenga pairs satisfying the assumptions of Conjecture~\ref{conj:logopeng} \cite{Bousseau:2020fus}. In particular, the fact that the higher genus log generating functions $\bbN_d^{\rm \log}(X,D)$ are integral Laurent polynomials (see e.g. \eqref{eq:p2hqg}), which is a consequence of $q$-deformed tropical correspondence argument, is refined here to an priori unexpected, stricter integrality statement.
\begin{theorem}
For all nef Looijenga pairs, we have that
\beq
 \left( \prod_{i=1}^{l} \frac{ 1 }{[ d \cdot D_i]_{q}} \right)
 \sum_{k | d}
\frac{(-1)^{ d/k \cdot D + l}\mu(k)}{[k]_{q}^{2-l} \,  k^{2-l} \,} \,
\bbN_{d/k}^{\rm log}(X,D)(q^k) \in \bbZ[q,q^{-1}]\,.
\label{eq:logbpsint}
\eeq
\end{theorem}
 The relation to BPS invariants echoes very similar\footnote{A non-trivial difference is that here the log Gromov--Witten invariants are {\it not} interpreted as BPS invariants themselves, unlike in \cites{Bou18,bousseau2018quantum}, but are instead related to them via \eqref{eq:logopeng} and \eqref{eq:lmovg}.}
statements relating log GW theory to DT and LMOV invariants in \cites{Bou18,bousseau2018quantum}, and in particular it partly demystifies the interpretation of log GW partition functions as related to some putative open curve counting theory on a Calabi--Yau 3-fold in \cite[\S 9]{bousseau2018quantum} by realizing the open BPS count in terms of actual, explicit special Lagrangians in a toric Calabi--Yau threefold.

\subsection{Further applications}
The open Gromov--Witten theory of toric Calabi--Yau 3-folds has been the subject of huge interest in the last couple of decades from many different communities. As a result, its identification with the higher genus log GW theory of surfaces brings with itself an immense range of implications for the latter, three of which were explored above (the topological vertex formalism being an effective means of closed-form resummation of the $q$-tropical counts; the integrality of KP invariants from the branes-quivers correspondence; and the open BPS integrality property of the log invariants as a result of the properties of LMOV partition functions).  The list does not end here, and we just highlight a few other connections that are immediate corollaries of Theorem~\ref{thm:logopeng}.
\bit
\item  The log invariants are naturally given by certain vacuum expectation values of a statistical mechanical model/topological quantum field theory: this can take the shape of either the resolvent of a random matrix ensemble \cites{Marino:2002fk,Brini:2011wi}, or as a melting crystal/free fermion vertex operator
\cites{Okounkov:2003sp,Saulina:2004da,MR2825318}, or yet again as Wilson loop in Chern--Simons theory \cites{Gopakumar:1998ki,Ooguri:1999bv});
\item Their generating functions are also related to $\tau$-functions of a classical integrable hierarchy, which is always a rational reduction of the 2-Toda hierarchy \cites{Brini:2010ap,Brini:2011ff,Brini:2014mha,Takasaki:2013gja,Takasaki:2013lqa},
\item Their genus expansion is computed by the Eynard--Orantin topological recursion, via the remodeled-B-model proposal
\cites{Marino:2006hs,Bouchard:2007ys,Eynard:2012nj,Fang:2016svw};
\item Finally for $l=2$, when they are related to disk superpotentials, their generating functions in genus zero are related to twisted superpotentials/vortex partition functions associated to surface operator insertion in a four-dimensional $\cN=2$ theory \cites{Kozcaz:2010af,Dimofte:2010tz}
\eit

The range of these fascinating implications are certainly worthy of further analysis, which we defer to future work.
\section{Generalizations}
\label{sec:gen}
The previous section saw a multitude of different curve counting theories (and corresponding physical theories) being non-trivially identified, starting from the datum of a nef Looijenga pair $(X,D)$. An obvious question is how rare is this web of correspondences -- how easy is it to build a pair $(X,D)$ with the required properties? In particular, in Definition~\ref{def:lp} we made no comments on how restrictive our assumptions are, especially the smoothness and nefness of each irreducible component $D_i$. It turns out that there are only finitely many (eighteen) smooth deformation families of nef Looijenga pairs, for which representatives will share the same Gromov--Witten invariants by deformation invariance. A key question is therefore to see how much the philosophy of Figure \ref{fig:web} carries through to as general a setup as possible.

\subsection{Orbifolds}
\label{sec:orbifolds}
\begin{figure}[t]
\centering
\begin{tikzpicture}[scale=2.4]
\node at (.2,0) {$\mathbb{C}^d$};
\draw [fill] (0,0) circle [radius=0.04];
\draw (-0.8,0) circle [x radius=0.8, y radius=0.6];
%
\draw[dashed] (-0.6,0) circle [x radius=0.6, y radius=0.45];
\draw[dashed] (-0.4,0) circle [x radius=0.4, y radius=0.3];
\draw (-0.2,0) circle [x radius=0.2, y radius=0.15];
\end{tikzpicture}
\caption{The $(n+1)$-loop quiver $\mathsf{Q}_{(n+1)-{\rm loop}}$ associated to $X=\bbC\bbP(1,1,n)$, $D=L_n+Q_n$.}
\label{fig:p1nnquiv}
\end{figure}
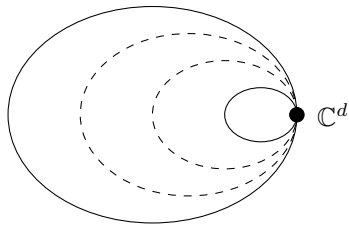

The first requirement we may want to drop is that the surface $X$ itself be smooth, and allow it to have orbifold (canonical quotient) singularities \cite{Bousseau:2019bwv,Bousseau:2020ryp}. In particular we can consider pairs $(\mathcal{X},D=D_1+\cdots +D_l)$ 
where $\mathcal{X}$ is a smooth complex Deligne--Mumford stack with coarse moduli space a normal Gorenstein projective surface $X$, $(X,D)$ is log-smooth (in particular, the singularities are concentrated along the codimension 2 strata of $D$), $D\in|-K_X|$, and the irreducible components $D_j$ are nef and $\bbQ$-Cartier for all $j=1, \dots, l$. The log-smoothness guarantees properness and existence of a virtual fundamental class for  the moduli space of basic stable log maps  \cites{Chen14,AbramChen14, GS13}, and the corresponding log Gromov--Witten invariants. Likewise, we may define local {\it orbifold} Gromov--Witten invariants of the non-compact Calabi--Yau orbifold $\mathcal{E}_{(X,D)} \coloneqq \mathrm{Tot}(\oplus_{i=1}^l (\cO_{\cX}(-D_i)))$ with coarse space the Gorenstein quasi-projective Calabi--Yau $(l+2)$-fold $E_{(X,D)}\coloneqq \mathrm{Tot}(\oplus_{i=1}^l (\cO_{X}(-D_i)))$ \cite{MR2450211}, and compute them using the orbifold version of the quantum Riemann--Roch theorem \cite{MR2578300}.
\begin{example}
Let $X=\bbC\bbP(1,1,n)$ be the weighted complex projective plane with weights $(1,1,n)$.  This is a Gorenstein toric surface which has one torus fixed point given by an orbifold singularity, which is locally a quotient of $\bbC^2$ by the finite cyclic group $\mu_n$. The singularity is joined to either of the other (smooth) torus-fixed points by a toric divisor $H_n$: extending  $H_n$ to an anticanonical divisor by adding a general member $Q_n$ of $|-K_Y-D_{1}|$ gives the Looijenga orbi-pair $(\bbC\bbP(1,1,n),D=H_n+Q_n)$. For $n=1$, this gives back our pet example $(X,D)=(\bbC\bbP^2, H \cup Q)$ in the smooth case.
\label{ex:p11n}
\vspace{.5cm}\end{example}
So by playing with the order of the singularities, we can generate an {\it infinite} list of nef Looijenga orbi-pairs, for which the log and local theories make sense. Moreover, and strikingly, the existence of an open string special Lagrangian pair $(Y,L)$ also carries through to this setting -- and with it,  the open BPS integrality statements of Section~\ref{sec:bpsdt}. In general, $(Y,L)$ may be an Aganagic--Vafa orbi-pair \cite{MR2861610,Brini:2013zsa}, and a slight refinement of the log-open correspondence may be required in that case -- we refer the reader to \cite{Bousseau:2020ryp} for a more extensive discussion.
\begin{example}
Let $(X,D)=(\bbC\bbP(1,1,n),H_n+Q_n)$ be as in Example~\ref{ex:p11n}. Note that $X$ is toric, $H_n$ is a prime toric divisor, and $Q_n$ is ample: this puts us squarely in the set of assumptions of Conjecture~\ref{conj:logopeng}. Running the same heuristic strategy of Section~\ref{sec:logopen}, we replace the divisor $H_n$ with a special Lagrangian condition near it, $Q_n$ with a twisting of $X \setminus H_n$ by $\cO(-Q_n)|_{X\setminus H_n}$, and then degenerate to a toric limit for the Lagrangian all the while remembering the datum of the compactification of $X \setminus H_n$ by adding back $H_n$ through a framing shift -- see Figure~\ref{fig:p11nop}. The associated threefold special Lagrangian pair in this case is given by $Y=\mathrm{Tot}(\cO(-Q_n)|_{X\setminus H_n})=\mathrm{Tot}(\cO_{\bbC\bbP(1,1,n)}(-Q_n)|_{\bbC^2}) \simeq \bbC^3$, and $L$ is an Aganagic--Vafa Lagrangian at framing $n$. In particular, although $X$ is singular, the pair $(Y,L)$ is smooth in this case. \\

\begin{figure}[!h]
\centering
\includegraphics[scale=0.75]{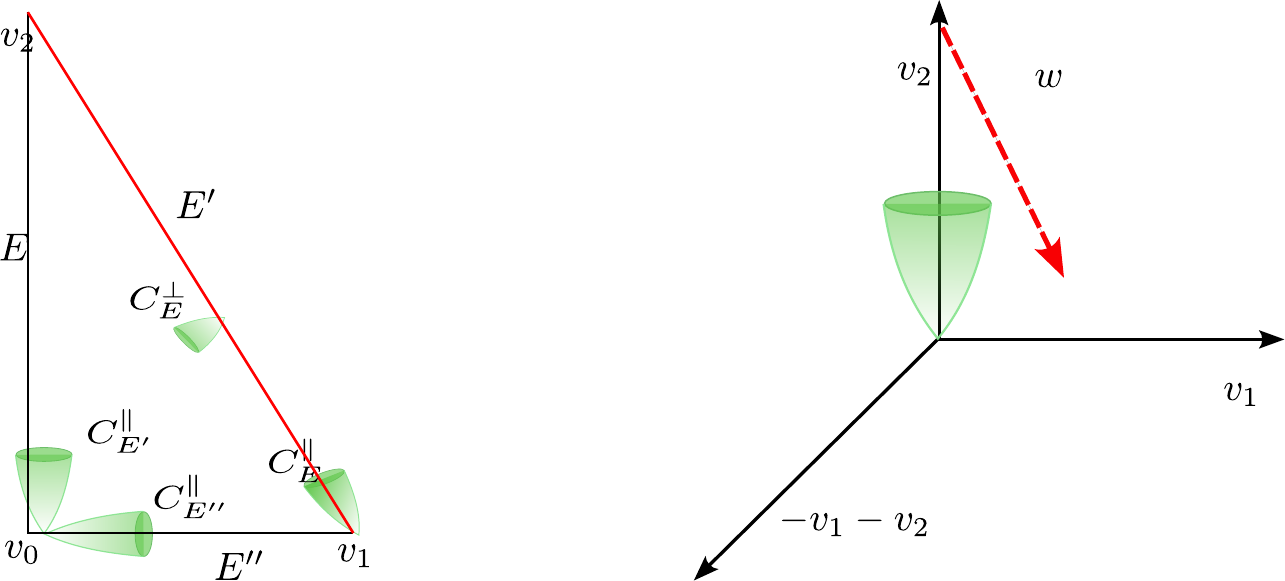}
\caption{The moment polytope of $\bbC\bbP(1,1,n)$ (left) for $n=2$, with a depiction of the generators $C_{E}^\|$ and $C_E^\perp$ (resp. $C_{E'}^\|$ and $C_{E''}^\|$) of the homology of the torus fiber $\ell$; and the corresponding toric diagram of $\bbC^3$ (right) with a toric brane on the edge $E'$ at framing $f=n$. The framing vector $w=v_1-f v_2= v_1-n v_2$ is parallel to $E$.}
\label{fig:p11nop}
\end{figure}

By \cite{Panfil:2018faz}, the associated quiver is the $(n+1)$-loop quiver (see Figure~\ref{fig:p1nnquiv}). For the corresponding invariants, the calculation of the open invariants is fundamentally identical to that of Example~\ref{ex:p2hqvert}, and we get
\beq
\bbO_{d}(Y,L) = \frac{(-1)^d}{d [(n+1) d]_q} \qbinom{(n+1) d}{d}_q
\label{eq:p1nnop}
\eeq
which agree with the $q$-scattering calculation \cite{Bousseau:2020ryp} of the higher genus log invariants of $(\bbC\bbP(1,1,n),H_n+Q_n)$ up to a factor of $(-1)^d d [(n+1) d]_q$, as predicted by Conjecture~\ref{conj:logopeng}:
\beq
\bbN^{\rm log}_{d}(\bbC\bbP(1,1,n),H_n+Q_n)= \qbinom{(n+1) d}{d}_q\,.
\eeq
Furthermore, \eqref{eq:p1nnop} gives, in the limit $q\to 1$, the local orbifold Gromov--Witten invariants of the orbifold CY4-fold geometry $E_{(\bbC\bbP(1,1,n),H_n+Q_n)}$. The (minimal) quiver associated to a single toric brane in $\bbC^3$ with framing $n$ is the $(n+1)$-loop vertex $\mathsf{Q}_{(n+1)-{\rm loop}}$ \cite{Panfil:2018faz}, see Figure~\ref{fig:p1nnquiv}. The corresponding numerical Donaldson--Thomas invariants are related to the Klemm--Pandharipande invariants of $E_{(\bbC\bbP(1,1,n),H_n+Q_n)}$ as
\bea
& & \mathrm{KP}_d \l(E_{(\bbC\bbP(1,1,n), H_n+Q_n)}\r) =
\bigg\{(-1)^n,\frac{1}{4}  \left( (2 n+1)-(-1)^n\right),\frac{1}{2} (-1)^n n (n+1),\nn \\ 
& & \frac{1}{3} n (n+1) (2 n+1),\frac{5}{24} (-1)^n n (n+1) (5 n
   (n+1)+2), \dots\bigg\}_d
   \nn \\
 & &  = (-1)^{n+d+1}\mathrm{DT}_d\l(\mathsf{Q}_{(n+1)-{\rm loop}}\r) \in \bbZ\,.
\eea 
\vspace{.5cm}\end{example}
\subsection{Non-nef and singular divisors}
\label{sec:non-nef}
A second, heavily constraining requirement in Definition~\ref{def:lp} was that each smooth component $D_i$, $i=1, \dots, l$ be nef. This places very strong constraints on $X$ and $D$: for example, when  $(X,D)$ is a toric pair, it forces $X$ to be a product of projective spaces. Recall that the nefness condition was chosen to ensure the generic compactness of  the moduli space of stable maps to the non-compact geometry $E_{(X,D)}$. But instead of imposing this as a condition on $D$, we can force that as a condition on our stable maps, by requiring that the class of their images lands in the nef cone of $X$. This vastly enlarges the catalogue of pairs $(X,D)$ amenable to the same analysis as the nef Looijenga pairs. 

Given a Looijenga pair $(X,D)$ there are two main birational operations that produce another Looijenga pair $(X',D')$  
\begin{itemize}
\item $X'$ is the blow-up of  $X$ at a node of $D$,  and $D'$ is the inverse image of $D$ in $X'$ (a \emph{corner blow-up} of $(X,D)$);
\item $X'$ is the blow-up of  $X$ at a smooth point of $D$,  and $D'$ is the strict transform of $D$ in $X'$ (an \emph{interior blow-up} of $(X,D)$);
\end{itemize}

A corner blow-up does not change the complement $X\setminus D$, whereas an interior blow-up does; accordingly corner blowups do not change log Gromov--Witten invariants \cite{AW}. By \cite{Fri}, every Looijenga pair $(X',D')$ dominates by a sequence of corner and interior blow-ups a \emph{minimal} Looijenga pair $(X,D)$ with $X$ a minimal rational surface. These can be classified, up to deformation, in four series, according to the number $l$ of irreducible components of $D$. 
\ben
\item for $l=1$, there are two isolated cases:
\bit
\item $X=\bbC\bbP^2$ and $D$ is an irreducible nodal cubic; 
\item $X=\bbC\bbP^1 \times \bbC \bbP^1$ and $D$ is a nodal bisection;
\eit
\item for $l=2$, there are three cases:
\bit
\item $X=\bbC\bbP^2$, $D=H \cup Q$;
\item $X=\bbF_n$, $n \neq 1$ is the $n^{\rm th}$ Hirzebruch surface, with $D_1= C_{-n}$ being the negative section and $D_2$ a smooth member of $|2 f + C_n|$, with $f$ the fiber class and $C_{n}$ the positive section; 
\item $X=\bbC\bbP^1 \times \bbC\bbP^1$ and $D_1=D_2$ is the class of the diagonal;
\eit
\item for $l=3$, there are two cases:
\bit
\item $X=\bbC\bbP^2$, $D=H \cup H \cup H$;
\item $X=\bbF_n$, $D_1= C_{-n}$, $D_2=f$, $D_3 \in |f+C_n|$;
\eit
\item for $l=4$, $X=\bbF_n$, $D$ is the toric boundary.
\een
The above examples with $X=\bbF_n$, $n>0$ are not nef, and we therefore have an infinite class of Looijenga pairs for each $l=2,3,4$. \\

Let's then restrict to stable map degrees in the nef cone of $X$, which is a full-dimensional subcone of the cone of effective curves. How do Theorems~\ref{thm:loglocal} and \ref{thm:logopeng} generalize to non-nef geometries? For $l>2$, it turns out that the explicit solution methods of the local, log, and open Gromov--Witten theories associated to $(X,D)$ presented in Sections~\ref{sec:calculation} and \ref{sec:vertex} can be applied seamlessly to the non-nef setting as well, and the resulting invariants satisfy the expected relations \eqref{eq:vggr}, \eqref{eq:locopen} and \eqref{eq:logopeng} \cite{bgns}. For $l=2$, the comparison argument presented in Section~\ref{sec:comparison} only relies on the fact that the stable map degree $d$ satisfies $d \cdot D_i >0$ for $i=1,2$, regardless of whether $D_i$ is nef, and it therefore holds unaltered in the non-nef setting, as does the comparison theorem of \cite{Liu:2021eeb,Liu:2022swc} for the local and open theory. Furthermore, in this case an all-degree calculation of the log and local invariants is computationally completely out of reach: because $D_1$ is not nef, the local mirror symmetry computations require a Birkhoff factorization of the $I$-function to extract the $J$-function, and the scattering calculation of the log invariants exhibit wall crossings with dense sets of walls in some sectors, thus making calculations in all degrees unfeasible. On the other hand, a topological vertex solution for the open theory can be determined in terms of the planar solution of a unitary matrix model \cite{Caporaso:2006gk,Eynard:2008mt}: the log-local-open comparison then produces, in a single shot an explicit algebraic formula for the generatic functions of the log, local, and open invariants in genus zero in terms of the planar resolvent of the chiral part of $q$-deformed two-dimensional Yang--Mills theory ($q\mathrm{YM}_2$) \cite{Aganagic:2004js, Caporaso:2005fp, Caporaso:2006gk,Marino:2006hs}. 
\begin{example} For the case of $\bbF_2$ with $D_1=C_{-2}$, $D_2 \in |2f+C_2|$ we have that $Y= \mathrm{Tot}(\cO_{\bbF_2}(-D_2)|_{\bbF_2 \setminus C_{-2}}) =  \mathrm{Tot}\l(\mathcal{O}_{\bbC\bbP^1}(2)\oplus \mathcal{O}_{\bbC\bbP^1}(-4)\r)$,  $L$ is a canonically framed special Lagrangian on the outer edge corresponding to the $\cO(2)$ fiber, and 
\beq
N_{0, d_0 C_2 + d_1 f}^{\rm log}(\bbF_2,C_{-2}+(2f+C_2)) = [Q^{d_0} z^{d_1}] (4 Q \partial_Q + z \partial_z) \log \phi(z,\zeta(Q))
\eeq
where  $\zeta(Q)$ is the unique root of $\zeta(Q)(1-\zeta(Q))^8 =Q$ which vanishes at $Q=0$, and
\beq
\phi(z,\zeta) \coloneqq
\frac{(1-\zeta )^{12}\left(\sqrt{\frac{1}{z}-\frac{1}{\left(\sqrt{\zeta }-1\right)^2 \left(\sqrt{\zeta }+1\right)^4}}+\sqrt{\frac{1}{z}-\frac{\left(\sqrt{\zeta }+1\right)^2}{(\zeta
   -1)^4}}\right)^{-2}}{ \left((\zeta -1)^3+\sqrt{(\zeta -1)^6+z^2-2 (\zeta +1) (\zeta -1)^2 z}-z\right)^4}\,.
\eeq
\begin{figure}[h]
\centering
\includegraphics[scale=0.9]{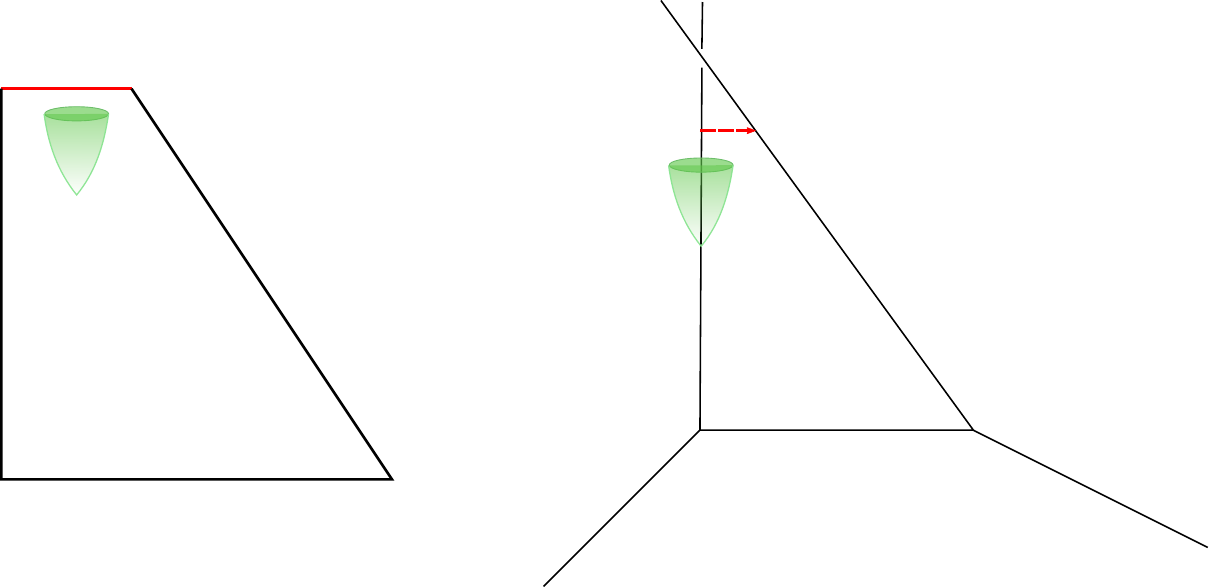}
\caption{The moment polytope of $\bbF_n$ (left) for $n=2$ with the divisor $D_1=C_{-n}$ depicted in red, and the corresponding toric diagram of $\mathrm{Tot}\l(\mathcal{O}_{\bbC\bbP^1}(n)\oplus \mathcal{O}_{\bbC\bbP^1}(-n-2)\r)$ (right), with a toric brane on an outer edge. Note that the toric diagram is non-planar, reflecting the non-nefness of $C_{-n}$.}
\label{fig:fn}
\end{figure}
\noindent Remarkably, not only the resolvent, but also the partition function of the unitary matrix model arising from $q\mathrm{YM}_2$ has an interpretation in terms of log and local Gromov--Witten counts. Upon relating log Gromov--Witten counts into $(\bbC\bbP^2, D_{\rm nodal})$ where $D_{\rm nodal}$ is a nodal cubic to counts in the complement $\bbC\bbP^2 \setminus {\rm pt} \simeq \mathrm{Tot}(\cO_{\bbC\bbP^1}(1))$ of the node, an all-genus comparison statement can be made in terms of the equivariantly Calabi--Yau Gromov--Witten theory of $E_{(\bbC\bbP^2\setminus {\rm pt},D_{\rm nodal} \setminus \rm pt)} \simeq {\rm Tot}(\cO_{\bbC\bbP^1}(1)\oplus \cO_{\bbC\bbP^1}(-3))$ (equivalently, the local Gromov--Witten theory of the quasi-projective surface given by the total space of $\cO_{\bbP^1}(1)$):
\beq
\bbN_{d}^{\rm log}(\bbC\bbP^2, D_{\rm nodal}) = (-1)^{3 d +1} [3 d]_q \bbN_{d}^{\rm loc}(\cO_{\bbP^1}(1)).
\eeq
The r.h.s. is then computed from the degree-expansion of the topological vertex partition function on the toric CY3 $E_{(\bbC\bbP^2\setminus {\rm pt},D_{\rm nodal} \setminus \rm pt)}$. The details will be presented in \cite{bgns}.
\vspace{.5cm}\end{example}

\section{Conclusion and outlook}

The results presented in this survey tie together several disconnected strands of development in the study of enumerative invariants of log CY surfaces and allied geometries. One particularly attractive spin-off of the discussion is the construction of a wide array of theoretical methods to determine the corresponding invariants in a unified way. We conclude here with a brief discussion of their relation to similar questions in enumerative geometry and physics, highlighting along the way some important avenues of future research.

\bit
\item The connection between log GW invariants to DT invariants of quivers and to open BPS invariants has appeared in previous related work \cites{Bou18,bousseau2018quantum}, which invoked a relation between log GW counts on surfaces with some putative open topological string on a CY3 and LMOV-type counts. This speculation is made fully explicit in our study of Looiejnga pairs through the lens of the log-open correspondence of Section~\ref{sec:logopen}. This bears an immediate consequence for the {\it local} invariants, by identifying KP invariants of local surfaces with quiver DT invariants, and proving algebro-geometrically the integrality of the former via the latter. This opens, {\it inter alia}, a glimpse of a possibly pathway to establish a Calabi-Yau 4-fold Gromov-Witten/Donaldson-Thomas correspondence statement as suggested in \cite{MR3861701, cao2019stable}, at least in the simplest case of local surfaces, which would be most interesting to further develop.
\item Our discussion in Section~\ref{sec:lp} imposed a set of rather stringent conditions ($X$ being a surface, $X$ and the irreducible components $D_i$ being smooth, nefness of $D_i$, maximal contact along $D_i$) which we only partly lifted in the description of the generalizations of Section~\ref{sec:gen}. It would be very interesting, for example to consider the the log/open correspondence beyond the context of maximal tangency: splitting the contact order across multiple points on $D_i$ would be mirrored, accordingly, 
by considering multiple Lagrangian boundary conditions near the same divisor $D_i$. The topological vertex computes these just as efficiently as the single-winding amplitudes, as would the ``remodeling'' technology of \cite{Bouchard:2007ys}. This expectation can already be put on firm grounds and seen to be satisfied in the basic case of a canonically framed Lagrangian on an outer edge of $\bbC^3$ and arbitrary windings, which would correspond to the log GW theory of $\bbC\bbP^1 \times \bbC$ relative to the toric boundary, with maximal contact along the zero fiber $[0:1] \times \bbC$ and arbitrary tangency along $[1:0] \times \bbC$. Given all that is known about the theory of the topological vertex, it would be fascinating to explore how much this could tell us about the log counts on $(X,D)$, and the construction of quantum SYZ mirrors as in \cite{bousseau2018quantum}.
\item
Recently \cite{cao2019stable,MR3861701,MR4460281}, a theoretical understanding of KP invariants was sought using sheaf-counting theories for Calabi--Yau $4$-folds \cite{MR3692967,OT20}, which have led to conjecture relations between genus $0$ KP invariants 
and stable pair invariants on Calabi--Yau 4-folds: their verification for local surfaces in \cite{Cao:2020hoy} relied on the solution of the Gromov--Witten/Klemm--Pandharipande side given by Theorem~\ref{thm:loglocal}. It would be extremely interesting to pin down the exact role of the appearance of the symmetric quiver $\mathsf{Q}(Y,L)$ in this context: for example, for $(X,D)=(\bbC\bbP^2, H \cup Q)$, the moduli space of representations of the corresponding quiver $\mathsf{Q}_{2-\rm loop}$ is isomorphic to the moduli space of rank-$d$ $\cO(1)$-twisted Higgs bundles
 on $\bbC\bbP^1$, which in turn is an open part of the moduli space of one-dimensional coherent sheaves on the CY4-fold local geometry $\mathrm{Tot}(\cO_{\bbC\bbP^2}(-1)\oplus \cO_{\bbC\bbP^2}(-2))$. It would be fascinating to hammer out a precise virtual comparison statement between the sheaves and the quiver perspective, and work out its implications for the corresponding D-brane realizations in string theory. 
\item Finally, the existence of a natural refinements separately for the DT theory of the quiver and the open BPS invariants raises the question of the nature of the possible refinement of the CY-4 fold DT invariants for the local surface geometries coming from Looijenga pairs with $l=2$, in terms of some putative refined version of DT theory for CY4-folds.
\eit

\section*{Acknowledgments}

It is a pleasure to thank my collaborators P.~Bousseau, M.~van~Garrel, N.~Nabijou and Y.~Sch\"uler for countless discussions on the topics of this survey, and for helpful comments on the manuscript. The presentation given here grew out of a series of seminar and conference talks hosted in person or online at Leeds, ICTP Trieste, Glasgow, Canterbury, Warwick 3CinG, Skoltech Moscow, Munich, Berkeley, Harvard, Nottingham, Leysin, Birmingham,  SISSA Trieste, Institut Mittag-Leffler, Sheffield and Oxford between September 2019 and November 2022, and I am grateful to the audiences who sat through them for their valuable feedback. My research was carried out under the auspices of the GNFM-INDAM and supported by the Engineering and Physical Sciences Research Council through the grant no.~EP/S003657/2.

\bibliographystyle{ws-ijmpa}

\bibliography{miabiblio}

\end{document}